\documentclass[a4paper,11pt]{article}
\pdfoutput=1
\usepackage{jheppub}
\usepackage[T1]{fontenc}

\usepackage{HEP} 

\graphicspath{{figures/}}

\title{\boldmath Jet charge identification in the \eeZqq{} process at \Z{} pole}

\author[a,b]{Hanhua Cui}
\author[c,d]{Mingrui Zhao}
\author[a,e]{Yuexin Wang}
\author[a,b]{Hao Liang}
\author[a,1]{Manqi Ruan\note{Corresponding author.}}

\affiliation[a]{Institute of High Energy Physics, Chinese Academy of Sciences, Beijing 100049, China}
\affiliation[b]{University of Chinese Academy of Science (UCAS), Beijing 100049, China}
\affiliation[c]{Science and Technology on Nuclear Data Laboratory, China Institute of Atomic Energy, Beijing, China}
\affiliation[d]{Neils Bohr Institute, University of Copenhagen, Copenhargen, Denmark}
\affiliation[e]{China Center of Advanced Science and Technology, Beijing 100190, China}

\emailAdd{cuihanhua@ihep.ac.cn}
\emailAdd{manqi.ruan@ihep.ac.cn}

\abstract{

Accurate jet charge identification is essential for precise electroweak and flavor measurements at the high-energy frontier.
We propose a novel method called the Leading Particle Jet Charge method (LPJC) to determine the jet charge based on information of the leading charged particle.
Tested on \Zbb{} and \Zcc{} samples at center-of-mass energy of $91.2\,$\GeV, the LPJC achieves an \etp{} of 20\%/9\% for \cbjet, respectively.
In combination with the Weighted Jet Charge method (WJC), we develop a Heavy Flavor Jet Charge method (HFJC), which achieves an \etp{} of 39\%/20\% for \cbjet, respectively.
This paper also discusses the dependencies between jet charge identification performance and the fragmentation process of heavy flavor jets, as well as critical detector performances.

}

\begin{document}
\maketitle
\flushbottom

{\color{black}
The modified content is marked in red.
}


\section{Introduction}
\label{sec:Introduction}

In the Standard Model (SM) of particle physics, quarks, and gluons carry color charge and are unable to propagate freely in spacetime due to the phenomenon of color confinement~\cite{HFLAV:2022pwe,PhysRevD.101.076020}. As a result, when these colored SM particles are produced in high-energy colliders, they fragment into sets of collimated final state particles, primarily hadrons~\cite{Elder:2018owb}, creating streams of particles known as jets. It is of great interest to distinguish the original species of colored SM particles within a jet, for instance, in measuring the properties of the Higgs at the high-energy frontier~\cite{Field:1977fa,Li:2023tcr,Wong:2023vpx}.

Quarks or gluons can be identified from the final state particles corresponding to the initial colored particle. 
Technically, within high-energy frontier experiments, quark jets can be differentiated from gluons at Large Hadron Collider (LHC) experiments~\cite{ATLAS:2017dfg,Elder:2017bkd,ATLAS:2017nma,Frye:2017yrw,Larkoski:2014pca,Gallicchio:2011xq}. 
Moreover, at both the LHC and future electron-positron (\ee{}) Higgs factories, \cbjet{} and light jets can be distinguished from each other using flavor tagging algorithms~\cite{Lee:2022kdn,Ilten:2017rbd,CMS:2017yer}. 
Finally, quarks and anti-quarks can be identified using jet charge identification algorithms at both LEP and LHC experiments.

Accurately measuring the jet charge is crucial for determining the asymmetry between the production of particles and their anti-particles in EW processes, as it directly affects the production rates of charged particles and their subsequent decay processes. 
The forward-backward asymmetry (\AFB) is a fundamental observable that provides insights into the underlying electroweak interactions~\cite{Yan:2021veo,Zhao:2022lyl,dEnterria:2018jsx,DELPHI:2004wzo,OPAL:2003pfe,ALEPH:2001pzx,OPAL:2002vqi,ALEPH:2001mdb,DELPHI:1994yxx,L3:1992fsb}. 
Similarly, the electroweak mixing angle (\swsq) is a fundamental parameter that characterizes the strength of the electroweak force~\cite{L3:1998jgx,Schael:1991nf}. 
Accurate measurement of jet charge is essential in determining \AFB{} and \swsq
Furthermore, in flavor physics, the time-dependent \CP{} measurement plays a crucial role in studying the violation of \CP{} symmetry in particle decays~\cite{Chen:2021ftn,dArgent:2021xcs,Bertolin:2021xli,Maccolini:2021rom}. 

The performance of heavy flavor jet charge identification is measured by \efficiency, \mromega, and \etp~\cite{Heinicke:2229990,Belle-II:2018jsg}, defined as

\begin{equation}
\label{equ:efficiency}
\epsilon_{\mathrm{tag}} = \frac{\mathrm{N_{selected}}}{\mathrm{N_{all}}}
\end{equation}

\begin{equation}
\label{equ:omega}
\omega = \frac{\mathrm{N_{wrong}}}{\mathrm{N_{selected}}}
\end{equation}

\begin{equation}
\label{equ:etp}
\epsilon_{\mathrm{eff}} = \epsilon_{\mathrm{tag}} \ (1-2 \omega)^2 = \epsilon_{\mathrm{tag}} \ r^2
\end{equation}
{\color{black}where $\mathrm{N_{all}}$ denotes the total number of input particles, $\mathrm{N_{selected}}$ represents the number of particles that have been tagged, $\mathrm{N_{wrong}}$ represents the number of particles that have been misidentified,} 
and $\epsilon_{\mathrm{tag}}$ represents the selection efficiency, which indicates the level of final state particle utilization. 
The $\omega$ denotes the misjudgment rate of all tagged samples.
This \mromega{} is usually converted into the dilution factor $r = 1-2 \omega$, which reflects the correct judgment rate. 
The dilution factor $r \  = \ 0$ indicates a fully diluted flavor (no possible distinction between \bcjet{} and \bbarcbarjet), whereas a dilution factor r = 1 indicates a perfectly tagged flavor,
The \etp{} indicates the total performance of jet charge tagging.
Physics measurements reliant on jet charge identification, such as \swsq, typically demonstrate accuracy proportional to $1/\sqrt{\epsilon_{\mathrm{eff}} N}$~\cite{Zhao:2022lyl}, where $N$ indicates the number of events observed.

Multiple jet charge identifications have been investigated from several collider experiments.
The LEP experiments employ the \cbjet{} charge for the determination of asymmetry \AFB{} and \swsq{} in the production of \bcquark{} pairs
~\cite{dEnterria:2018jsx,DELPHI:2004wzo,OPAL:2003pfe,ALEPH:2001pzx,OPAL:2002vqi,ALEPH:2001mdb,CDF:1999jfn,LEPHeavyFlavorWorkingGroup:1998roh,L3:1998jgx,DELPHI:1994yxx,OPAL:1994xvz,L3:1992fsb,ALEPH:1991fba}
and for neutral \Bmeson{} oscillation studies~\cite{OPAL:1994xvz,CDF:1999jfn}.
Five methods have been used in LEP experiments for the jet charge identification, relying on leptons, \chadron s, momentum-weighted jet charge, vertex charge, and kaons.
DELPHI achieves a b purity of 90\% in double tagged events~\cite{DELPHI:2004wzo}.
The ATLAS and CMS Collaborations at the LHC use dijet events to differentiate between quark jets and gluon jets to test different aspects of the strong interaction~\cite{Bielcikova:2021ujk,CMS:2020plq,CMS:2017yer,ATLAS:2015rlw,Nachman:2014qma,Krohn:2012fg}, and use muon and momentum-weighted jet charge to measure \Bso{} decay parameters, achieving an $\epsilon_{\mathrm{eff}}$ of 1.49\%~\cite{ATLAS:2016pno}.
LHCb uses the b-tagging algorithm for precision measurement of \CP{} violation via neutral \Bmeson{} decay channels, using opposite side $e, \mu, K$, \chadron, vertex charge, same side $\pi, p, K$ ~\cite{LHCb:2018roe, Heinicke:2229990}, and combined with Quantum ML, reaching an \etp{} of about 8\% ({\color{black}transverse momentum \pT>60~\GeV })~\cite{Gianelle:2022unu}.
The concept of jet charge identification can be extended to heavy flavor factories, such as Belle II and BABAR.
These factories distinguish between \bquark{} and \bbarquark{} for time-dependent \CP{} analysis using hadronic $B$ decays with flavor-specific final states at the \ensuremath{\Upsilon(\mathrm{4S})} resonance.
With a combination of different tagging signatures and neural networks, 
BABAR achieves the an \etp{} of 31.2\%~\cite{BaBar:2009byl, BaBar:2012fgk}, 
while Belle II obtains the \etp{} of 30.0\%~\cite{Belle-II:2021zvj, Belle-II:2018jsg}.

The \ee{} Higgs factory has been identified as the foremost priority for future high energy particle collider experiments~\cite{EuropeanStrategyforParticlePhysicsPreparatoryGroup:2019qin}. 
Several facilities have been proposed,
including
CEPC~\cite{CEPCStudyGroup:2018rmc,CEPCStudyGroup:2018ghi}, FCC-$ee$~\cite{FCC:2018byv,FCC:2018evy,FCC:2018vvp}, ILC~\cite{ILC:2013jhg}, CLIC~\cite{Linssen:2012hp}, C$^3$~\cite{Bai:2021rdg}, and Helen~\cite{Belomestnykh:2022wbn}.
These multipurpose factories not only allow for precision Higgs property measurements, but also offer considerable potential for accurate measurements of electroweak (EW), quantum chromodynamics (QCD), flavor physics, and searches for new physics.
Especially, the circular \ee{} Higgs factories, such as the CEPC and the FCC-$ee$, could deliver Teras of \Z{} bosons and provide excellent opportunities for those explorations.
Accurate jet charge identification has a significant impact on the scientific reach of these \ee{} Higgs factories. 

In this study, we investigate the jet charge identification performance at the future \ee{} Higgs factory.
{\color{black}The influence of light flavors has been neglected because only 0.051/0.005 light flavors could be misidentified as \cbjet, respectively~\cite{Zhu:2023xpk}.}
We generate approximately 10 million \Zbb{} and \Zcc{} events using three distinct generators for comparison: \Whizard~\cite{Stienemeier:2021cse, Brass:2019hvu, Kilian:2007gr}, \Herwig~\cite{Bellm:2015jjp}, and \Sherpa~\cite{Sherpa:2019gpd}.
In \Whizard, the built-in \textsc{Pythia~6} \cite{Sjostrand:2006za} is used for jet fragmentation and hadronization.
Uncertainty of \etp{} raised from finite statistics of Monte Carlo (MC) samples, being $10^{-4}$ order, is negligible and thus omitted in this study.
We have developed an algorithm, LPJC, to identify the  charge of a single jet utilizing information from the final state leading charged particle within it.
Combing with the conventional method that uses the sum of jet particles' charge weighted by the energy proposed by~\cite{Waalewijn:2012sv}, we developed a Heavy Flavor Jet Charge method (HFJC), achieving an \etp{} of 39\%/20\% for \cbjet, respectively.
The \etp{} can be significantly improved to 45\%/37\% for \cbjet, respectively, by additionally identifying the origin of the final state leading charged particles. 

This paper is organized as follows.
Section~\ref{sec:LPJC} describes LPJC and its performance, and a brief analysis for the performance dependency on jet hadronization.
Section~\ref{sec:WJC} presents the corresponding information of WJC method.
We develop a Heavy Flavor Jet Charge method (HFJC) which combines WJC and LPJC methods.
Section~\ref{sec:Combination} introduces the combination method and HFJC performance, which conclude the characteristic jet charge performance at the \Z{} pole operation.
{\color{black}In Section~\ref{sec:Introduction}-~\ref{sec:Combination}, our assumptions regarding the detector are ideal, while in Section~\ref{sec:Detector}, we discuss the impact of actual detector performance and quantifies the variation of \etp{} at different conditions.}
A short summary and a discussion on perspectives are given in Section~\ref{sec:Summary}. 

\section{Leading Particle Jet Charge method (LPJC)}
\label{sec:LPJC}

Generated in high energy colliders, \bcquark s typically fragment into \bchadron s (hadrons that contain \bcquark), comprising both excited and ground states.
This study designates the most energetic ground-state heavy hadron within a jet as the leading heavy hadron.
Specifically, in \cjet s, these include \Dp, \Do, \Dsp, \Lambdacp, while in \bjet s, these comprise \Bobar, \Bn, \Bsobar, \Lambdabo, which collectively account for 99.7\% and 99.8\% of all heavy hadrons,  respectively.
The mass, lifetime $\tau$, decay length $c\tau$ and proportion of these hadrons are summarized in Table~\ref{tab:OH}.
According to \Whizard, approximately 67\% and 83\% of the final state leading charged particles in \cbjet, 
respectively, are from leading heavy hadron decays, which have a sizeable distance from the interaction point (IP). 
Apart from these, final state charged particles can also originate in proximity of IP, primarily through QCD fragmentation.
For the sake of simplicity, these origins in proximity of IP are collectively referred to as QCD fragmentation. 

\begin{table}[htb]
    \centering
    \renewcommand{\arraystretch}{1.2}
	\caption{The mass, lifetime $\tau$, decay length $c\tau$, and proportion for each type of leading heavy hadrons~\cite{ParticleDataGroup:2020ssz}.}
    \begin{tabular}[t]{|c|c|c|c|c|c|}
        \hline
        \multicolumn{2}{|c|}{\makecell{leading \\ heavy hadron}} & \makecell{mass \\ \MeV} & \makecell{$\tau$ \\ $ps$} & \makecell{$c\tau$ \\ $\mu m$} & proportion \\
        \hline
        \multirow{5}{*}{\cjet} & \Dp & 1869.66 & 1.033 & 309.8 & $22.0\%$ \\
        \cline{2-6}
        & \Do & 1864.84 & 41.00 & 123.0 & $62.8\%$ \\
        \cline{2-6}
        & \Dsp & 1968.35 & 50.40 & 151.2 & $8.1\%$ \\
        \cline{2-6}
        & \Lambdacp & 2286.46 & 20.15 & 60.4 & $6.8\%$ \\
        \cline{2-6}
        & Sum &  \multicolumn{3}{c|}{}  & $99.7\%$ \\
        \hline
        \multirow{5}{*}{\bjet} & \Bobar & 5279.66 & 1.519 & 455.4 & $42.4\%$ \\
        \cline{2-6}
        & \Bn & 5279.34 & 1.638 & 491.1 & $42.4\%$ \\
        \cline{2-6}
        & \Bsobar & 5366.92 & 1.520 & 455.7 & $8.2\%$ \\
        \cline{2-6}
        & \Lambdabo & 5619.60 & 1.471 & 441.0 & $6.8\%$ \\
        \cline{2-6}
        & Sum &  \multicolumn{3}{|c|}{}  & $99.8\%$ \\
        \hline
    \end{tabular}
    \label{tab:OH}
\end{table}

\begin{figure}[htbp]
  \centering
  \includegraphics[width=0.75\textwidth]{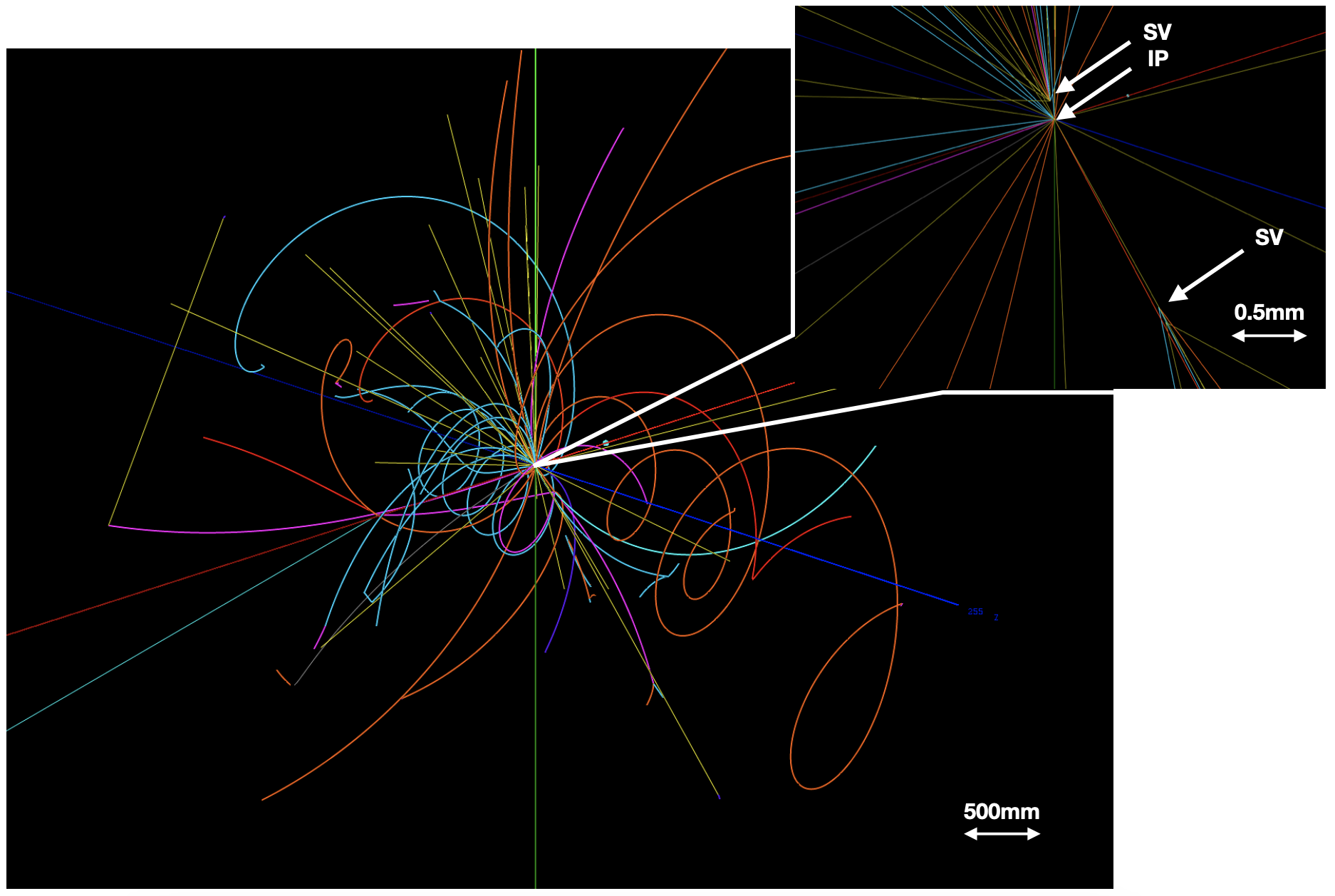}
  \caption{The jet event display of \eeZbb{} event at \Z{} pole generated by \Whizard.}
  \label{fig:display}
\end{figure}

Figure~\ref{fig:display} shows an display of a \Zbb{} event where the trajectories of charged particles are depicted by curves. 
All particles within the event can be clustered into two back-to-back jets based on the plane perpendicular to the thrust axis~\cite{Chang:2013iba}.
The enlarged view of the event display shows the IP and the secondary vertices (SV), 
which can be used to distinguish between particles originating from the leading heavy hadron decays and those produced via QCD fragmentation. 
A heavy flavor jet can be regarded as a composition of one leading heavy hadron along with a set of light hadrons, predominantly pions.

\begin{figure}[htbp]
  \centering
  \includegraphics[width=0.4\textwidth]{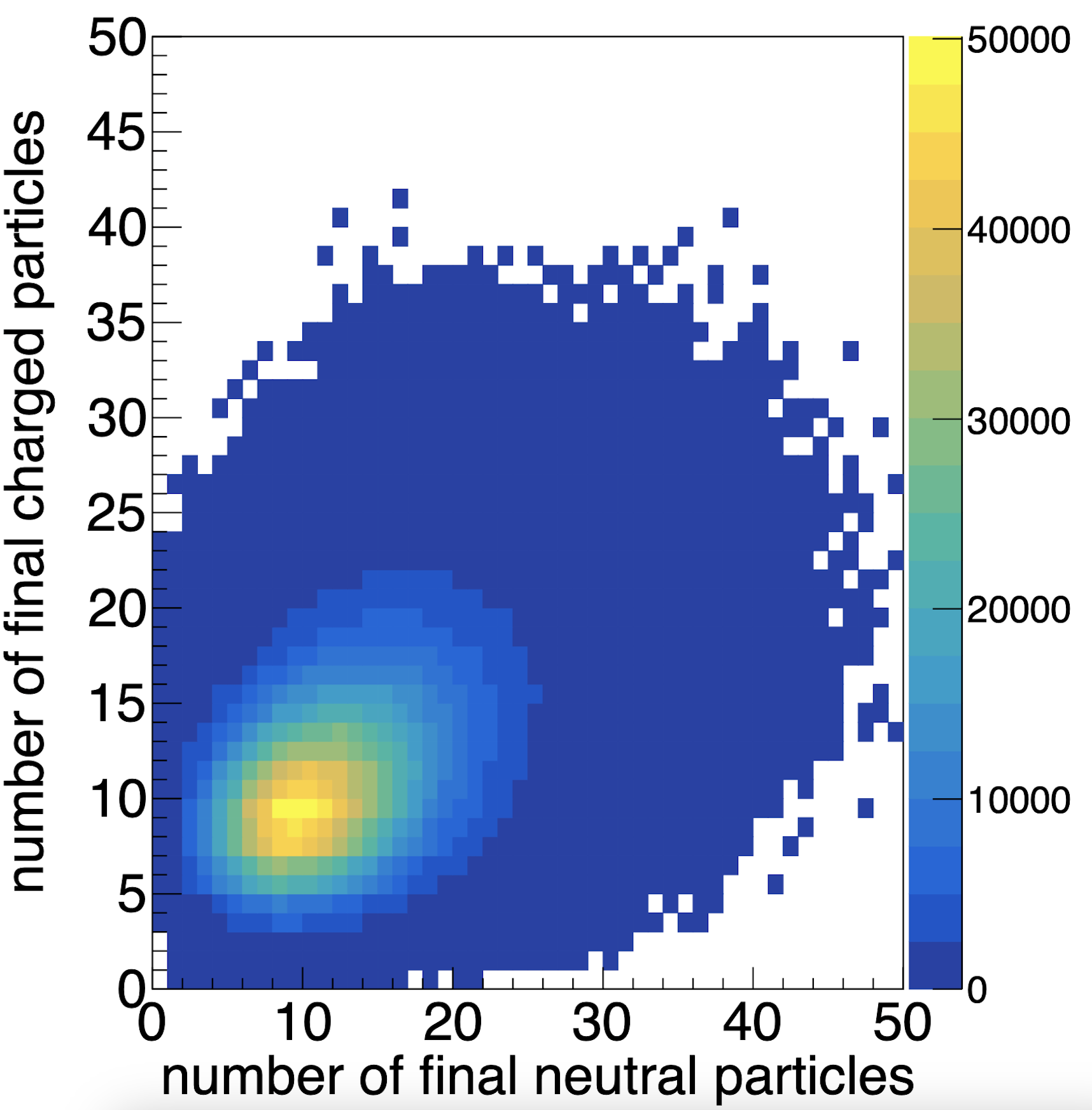}
  \hspace{0.8cm}
  \includegraphics[width=0.4\textwidth]{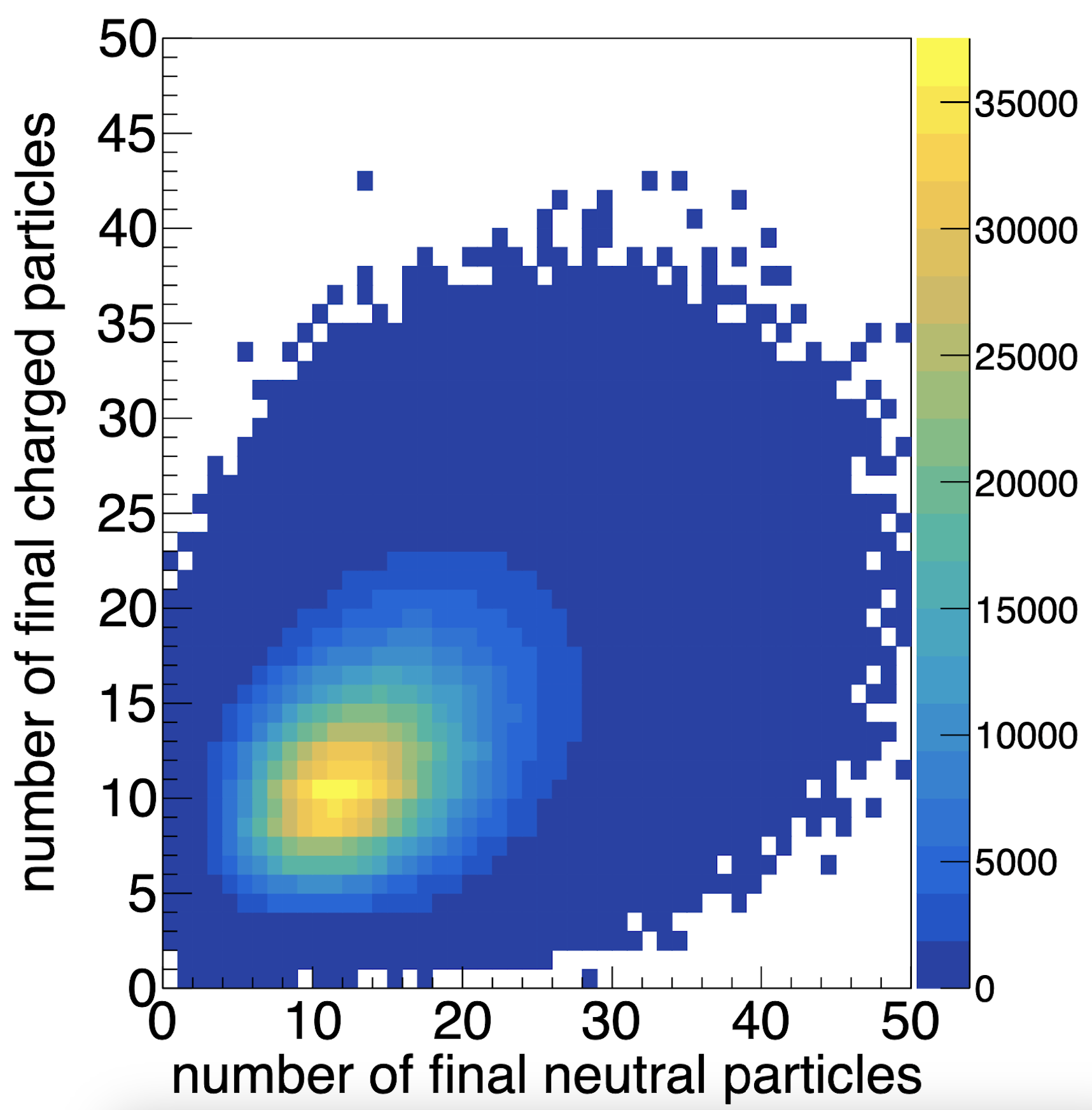}
  \caption{The multiplicity distributions of final state charged particles vs. neutral particles of \cjet  ~(left), and \bjet ~(right) on the \Z{} pole.}
  \label{fig:multiplicity}
\end{figure}

The jet charge identification algorithm takes the information of final state particles as input and determines the charge of the initial heavy quark.
The observable properties of the jet, such as momentum flow, multiplicity, and charge, reflect the characteristics of the heavy quarks. 
Figure~\ref{fig:multiplicity} illustrates the multiplicity of final state charged and neutral particles in \cbjet s,
which peaks at a value of 10 and exhibits an extended tail towards the larger values.
This section introduces the LPJC method and evaluates its performance in various decay processes.

\subsection{Methodology of LPJC}
\label{sec:LPJC Algorithm}

In this section, we introduce the LPJC algorithm and its performance quantified by the \etp.
The LPJC method is composed of three key steps: 
First, all final state particles in each event are clustered into two back-to-back jets. 
{\color{black}Secondly, select the charged particle with highest energy in each jet, identified as the final state leading charged particle.}
These are then classified into sub-groups based on their types: \emuKpipr. 
This covers almost all final state charged particles in \cbjet s.
Finally, the jet charge is determined using the charge and particle identification (PID) information of the final state leading charged particle in each sub-group, and the asymmetry between the two charged particles in one sub-group is defined as the corresponding \mromega.

\begin{figure}[htbp]
  \centering
  \includegraphics[width=0.4\textwidth]{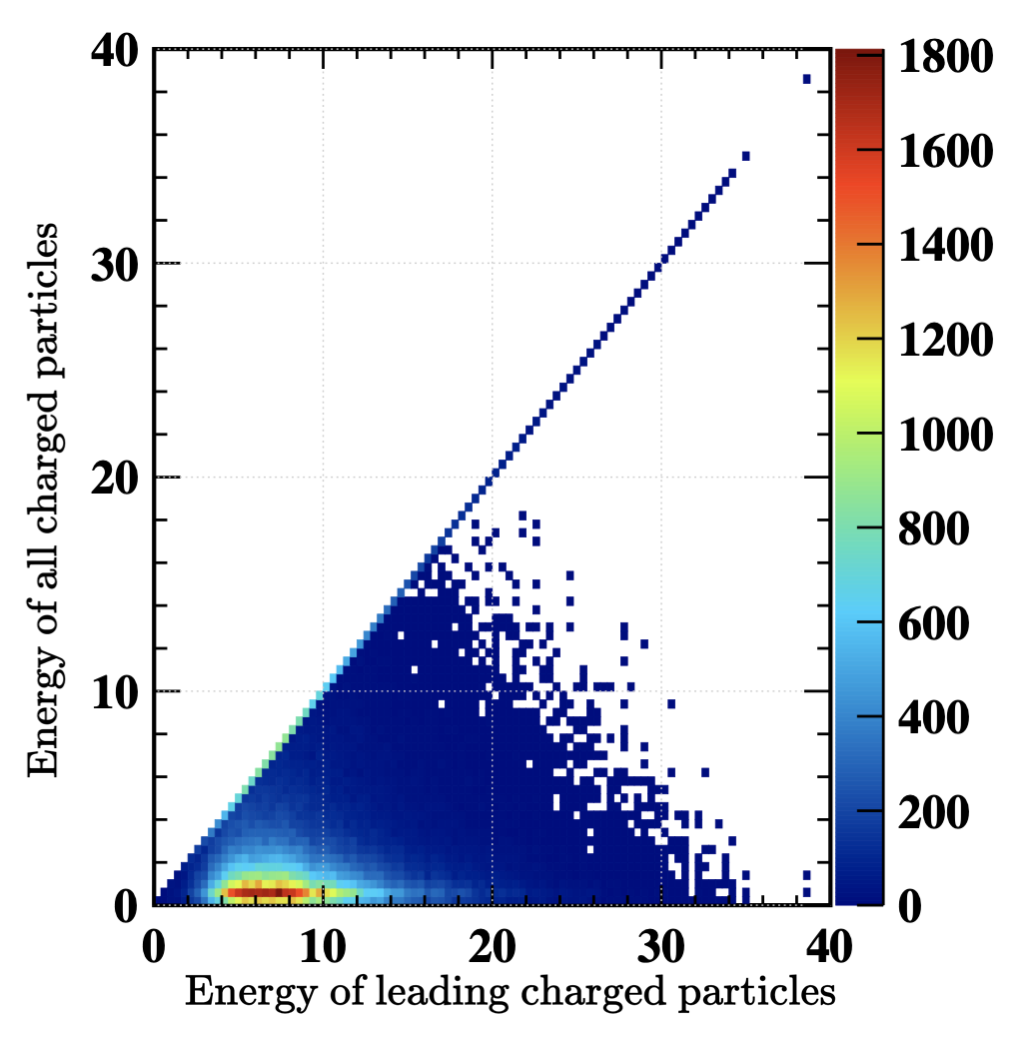}
  \includegraphics[width=0.4\textwidth]{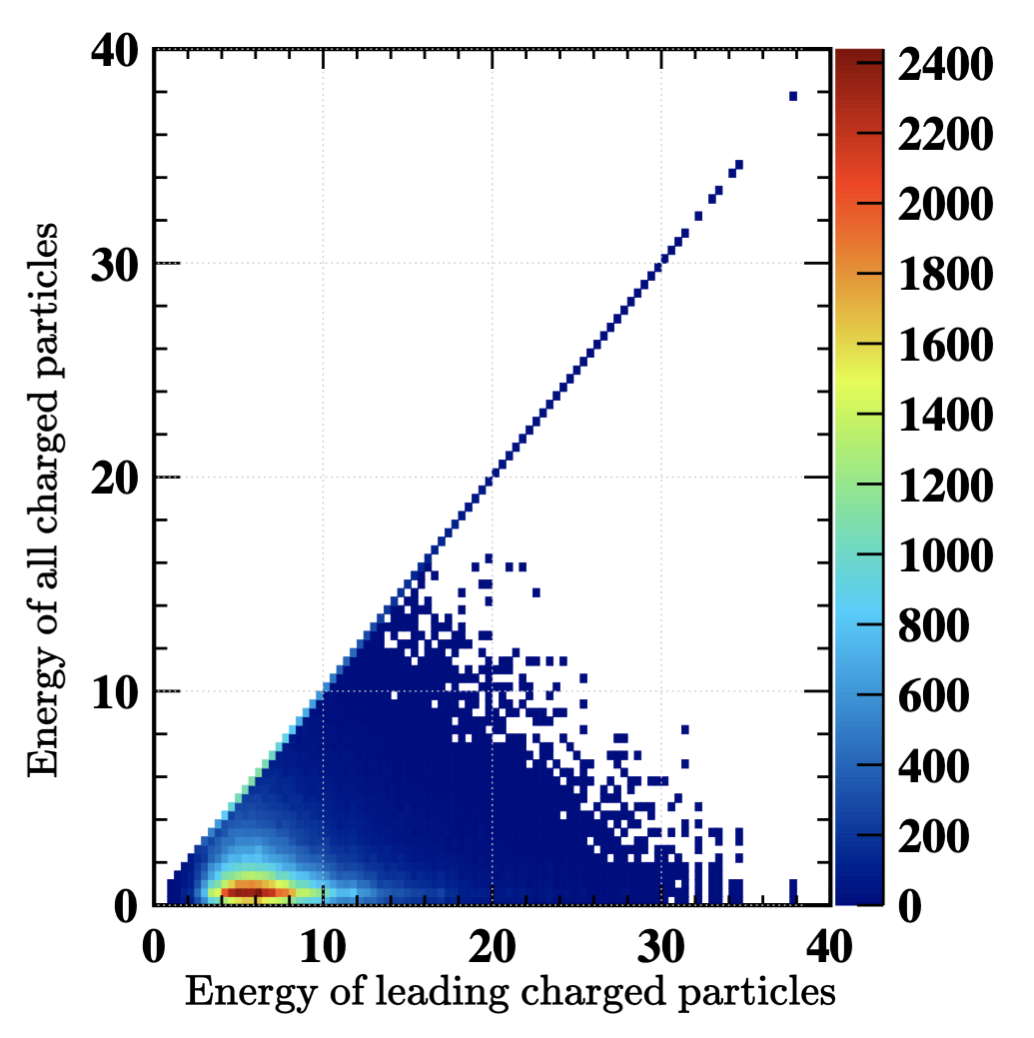}
  \caption{The energy spectrum of final state charged particles for \cjet{} (left) and \bjet{} (right) {\color{black} The entries along the diagonal represent the final state leading charged particles.}}
  \label{fig:Energy}
\end{figure}

{\color{black}The energy spectrum of final state charged particles is shown in figure~\ref{fig:Energy}.
The x-axis represents the energy of final state leading charged particles, ranging from 0 to 35 \GeV, with a peak of 5 to 10 \GeV. 
The y-axis represents the energy of all final state charged particles.
The entries along the diagonal represent the final state leading charged particles.}

\begin{figure}[htbp]
  \centering
  \includegraphics[width=\textwidth]{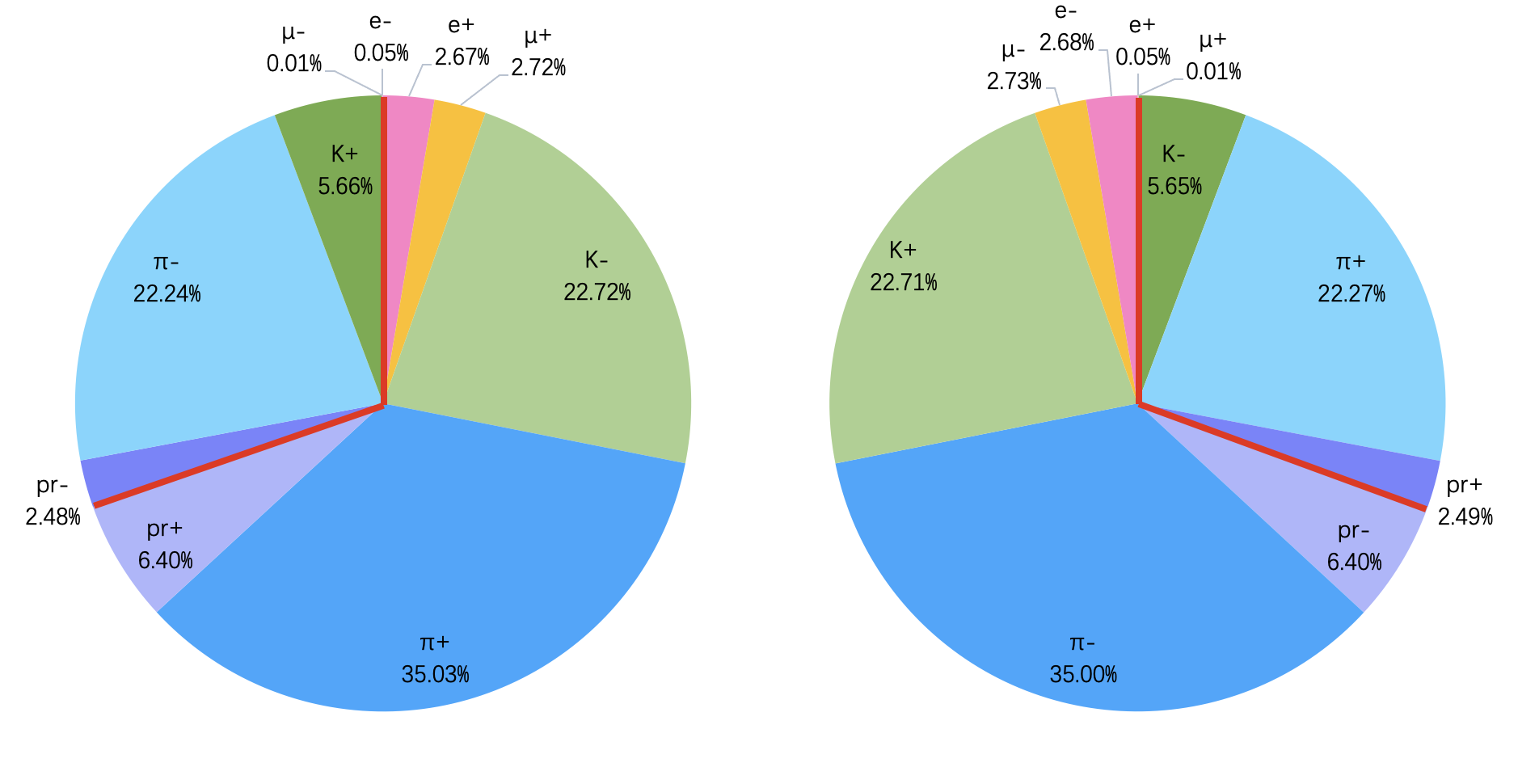}
  \includegraphics[width=0.95\textwidth]{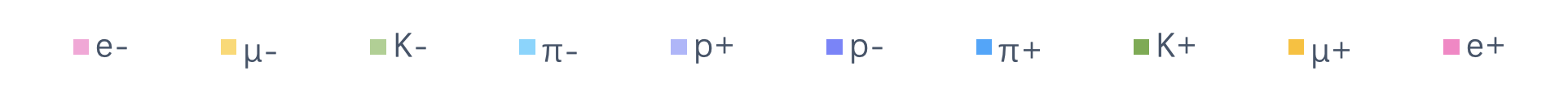}
  \caption{The percentages of species of final state leading charged particles within the \cjet{} (left) and \cbarjet{} (right) {\color{black}by \Whizard}.}
  \label{fig:Zcc_FP_distribution}
\end{figure}

\begin{figure}[htbp]
  \centering
  \includegraphics[width=\textwidth]{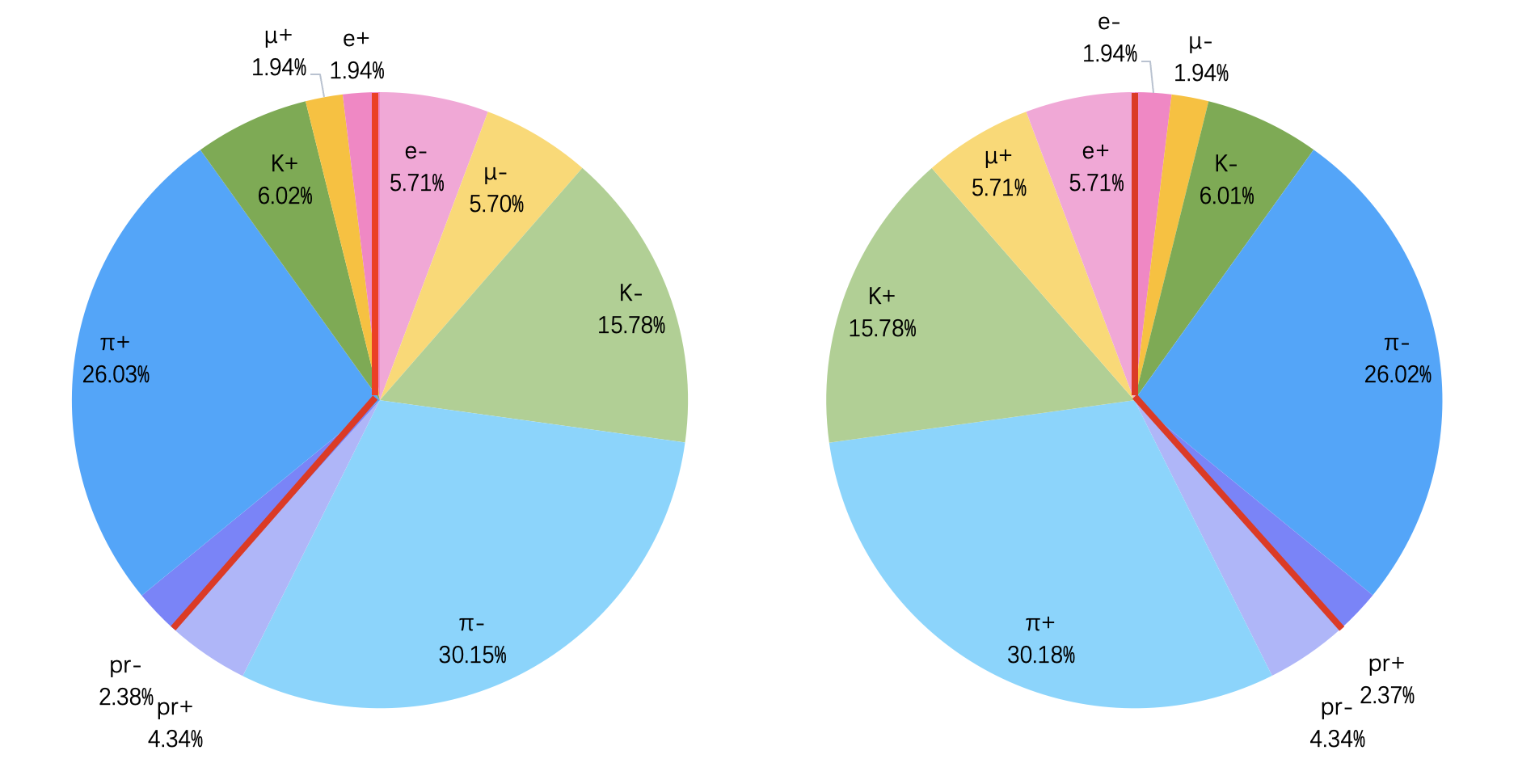}
  \includegraphics[width=0.95\textwidth]{./figures/distributions/leg_LP.png}
  \caption{The percentages of species of final state leading charged particles within the \bjet{} (left) and the \bbarjet{} (right) {\color{black}by \Whizard}.}
  \label{fig:Zbb_FP_distribution}
\end{figure}

\begin{table}[htb]
    \centering
    \renewcommand{\arraystretch}{1.2}
	\caption{The LPJC \etp{} of each type of final state leading charged particles and the total \etp{} at the \cbjet, which is the sum of each due to the independence.
	The \mromega{} for the sum are computed from the total \etp{} using eq. \eqref{equ:etp}.
	}
    \begin{tabular}[t]{|c|c|c|c|c|c|c|}
        \hline
        & & \multicolumn{3}{c|}{\Whizard} & \Herwig & \Sherpa \\
        \hline
        & & {\color{black}$\epsilon_{\mathrm{tag,max}}$} & $\omega$ & $\epsilon_{\mathrm{eff}}$ & $\epsilon_{\mathrm{eff}}$ & $\epsilon_{\mathrm{eff}}$ \\
        \hline
        \multirow{6}{*}{\cjet} & $e$ & $2.7\%$ & $1.9\%$ & $2.5\%$ & $3.4\%$ & $2.9\%$ \\
        \cline{2-7}
        & $\mu$ & $2.7\%$ & $0.5\%$ & $2.7\%$ & $3.0\%$ & $3.2\%$ \\
        \cline{2-7}
        & $K$ & $28.4\%$ & $19.7\%$ & $10.4\%$ & $11.8\%$ & $11.0\%$ \\
        \cline{2-7}
        & $\pi$ & $57.3\%$ & $38.8\%$ & $2.9\%$ & $3.4\%$ & $2.8\%$ \\
        \cline{2-7}
        & $p$ & $8.9\%$ & $28.0\%$ & $1.7\%$ & $0.9\%$ & $1.5\%$ \\
        \cline{2-7}
        & Sum & $100\%$ & $27.5\%$ & $20.2\%$ & $22.5\%$ & $21.5\%$ \\
        \hline
        \multirow{6}{*}{\bjet} & $e$ & $7.6\%$ & $25.5\%$ & $1.8\%$ & $1.8\%$ & $1.5\%$ \\
        \cline{2-7}
        & $\mu$ & $7.6\%$ & $25.5\%$ & $1.8\%$ & $2.1\%$ & $1.5\%$ \\
        \cline{2-7}
        & $K$ & $21.8\%$ & $27.5\%$ & $4.4\%$ & $3.3\%$ & $3.6\%$ \\
        \cline{2-7}
        & $\pi$ & $56.2\%$ & $46.3\%$ & $0.3\%$ & $0.5\%$ & $0.6\%$ \\
        \cline{2-7}
        & $p$ & $6.7\%$ & $36.5\%$ & $0.5\%$ & $0.7\%$ & $0.6\%$ \\
        \cline{2-7}
        & Sum & $100\%$ & $35.1\%$ & $8.9\%$ & $8.4\%$ & $7.8\%$ \\
        \hline
    \end{tabular} 
    \label{tab:ETP_LPJC}
\end{table}

{\color{black}The LPJC criteria for determining whether it is \cjet{} and \cbarjet{} (similarly for \bjet{} and \bbarjet) is the characteristics, including type and charge, of the final state leading charged particle. 
The specific correlation between the initial quark charge and the characteristics of the final state leading charged particle is outlined as follows:}

The electric charge of each final state leading charged particle exhibits an obvious asymmetry that reflects the jet charge \mromega, which can be interpreted in eq.~\eqref{equ:omega}.
Figure~\ref{fig:Zcc_FP_distribution} and~\ref{fig:Zbb_FP_distribution} show this asymmetry, with the red lines dividing the pies into two parts that indicate correct judgment ($N_\mathrm{R}$) and misjudgment ($N_\mathrm{W}$) rates. 
The charge sign for each species of final state leading charged particle exhibits an obvious asymmetry between \cjet{} and \cbarjet, which can be interpreted as that of the quark.
Therefore, the misjudgment rate $\omega$ can be calculated accordingly.
The \etp, calculated according to eq.~\eqref{equ:etp}, accounts for the effective reduction of events due to flavor dilution and represents the overall performance of the jet charge.

Table~\ref{tab:ETP_LPJC} provides the \etp{} values for each species of final state leading charged particle. 
{\color{black}The $\epsilon_{\mathrm{tag,max}}$ represents the upper limit on the tagging efficiency $\epsilon_{\mathrm{tag}}$.
The efficiency values in Table~\ref{tab:ETP_LPJC} are used for the subsequent calculations.}
The different species of final state leading charged particles form an un-overlapping, almost full coverage of all events, enabling the calculation of total \etp{} as a straightforward sum of each category's \etp~\cite{Heinicke:2229990, Belle-II:2018jsg}.
The \mromega{} for the sum is then computed from the total \etp{} using eq. \eqref{equ:etp}.

The LPJC method effectively utilizes final state leading charged particles to determine the jet charge, establishing benchmark values for the \etp.
For \bjet, these values are 8.9\% by \Whizard, 8.4\% by \Herwig, and 7.8\% by \Sherpa.
Meanwhile, for \cjet, these values are 20.2\% by \Whizard, 22.5\% by \Herwig, and 21.5\% by \Sherpa.

\subsection{Jet charge performance dependencies}
\label{sec:Dependencies}

The dependence of the correlation between final state particles and jet charge is not solely affected by the species and charge of final state leading charged particles but also by their origin: the final state leading charged particles can be generated from leading heavy hadron decay or QCD fragmentation. 
While for the former, the correlation will also be sensitive to the species of leading heavy hadron. 
This session analyzes the dependence of \mromega (the charge asymmetry of final state leading charged particles) in different categories of one jet, including:

\begin{enumerate}
\item The type of the final state leading charged particle;
\item The type of the leading heavy hadrons;
\item The sources of the final state leading charged particles, whether the final state leading charged particles are from the leading heavy hadron decay or from QCD fragmentation.
\end{enumerate} 

\begin{table}[htb]
    \centering
    \renewcommand{\arraystretch}{1.2}
	\caption{The \etp{} for each type of the leading heavy hadrons and the total \etp{} {\color{black}by \Whizard}.}
    \begin{tabular}[t]{|c|c|c|c|c|}
        \hline
        \multicolumn{2}{|c|}{leading hadron} & proportion & $\omega$ & $\epsilon_{\mathrm{eff}}$ \\
        \hline
        \multirow{5}{*}{\cjet} & \Dp & $22.0\%$ & $23.9\%$ & $19.3\%$ \\
        \cline{2-5}
        & \Do & $62.8\%$ & $19.0\%$ & $23.5\%$ \\
        \cline{2-5}
        & \Dsp & $8.1\%$ & $22.2\%$ & $14.1\%$ \\
        \cline{2-5}
        & \Lambdacp & $6.8\%$ & $20.0\%$ & $28.1\%$ \\
        \cline{2-5}
        & total & $99.7\%$ & $20.3\%$ & $22.1\%$ \\
        \hline
        \multirow{5}{*}{\bjet} & \Bobar & $42.4\%$ & $38.2\%$ & $5.6\%$ \\
        \cline{2-5}
        & \Bn & $42.4\%$ & $30.2\%$ & $15.8\%$ \\
        \cline{2-5}
        & \Bsobar & $8.2\%$ & $47.0\%$ & $0.4\%$ \\
        \cline{2-5}
        & \Lambdabo & $6.8\%$ & $23.1\%$ & $28.9\%$ \\
        \cline{2-5}
        & total & $99.8\%$ & $34.5\%$ & $11.0\%$ \\
        \hline
    \end{tabular}
    \label{tab:ETP_LPJC_OH}
\end{table}

\begin{figure}[htbp]
  \centering
  \includegraphics[width=\textwidth]{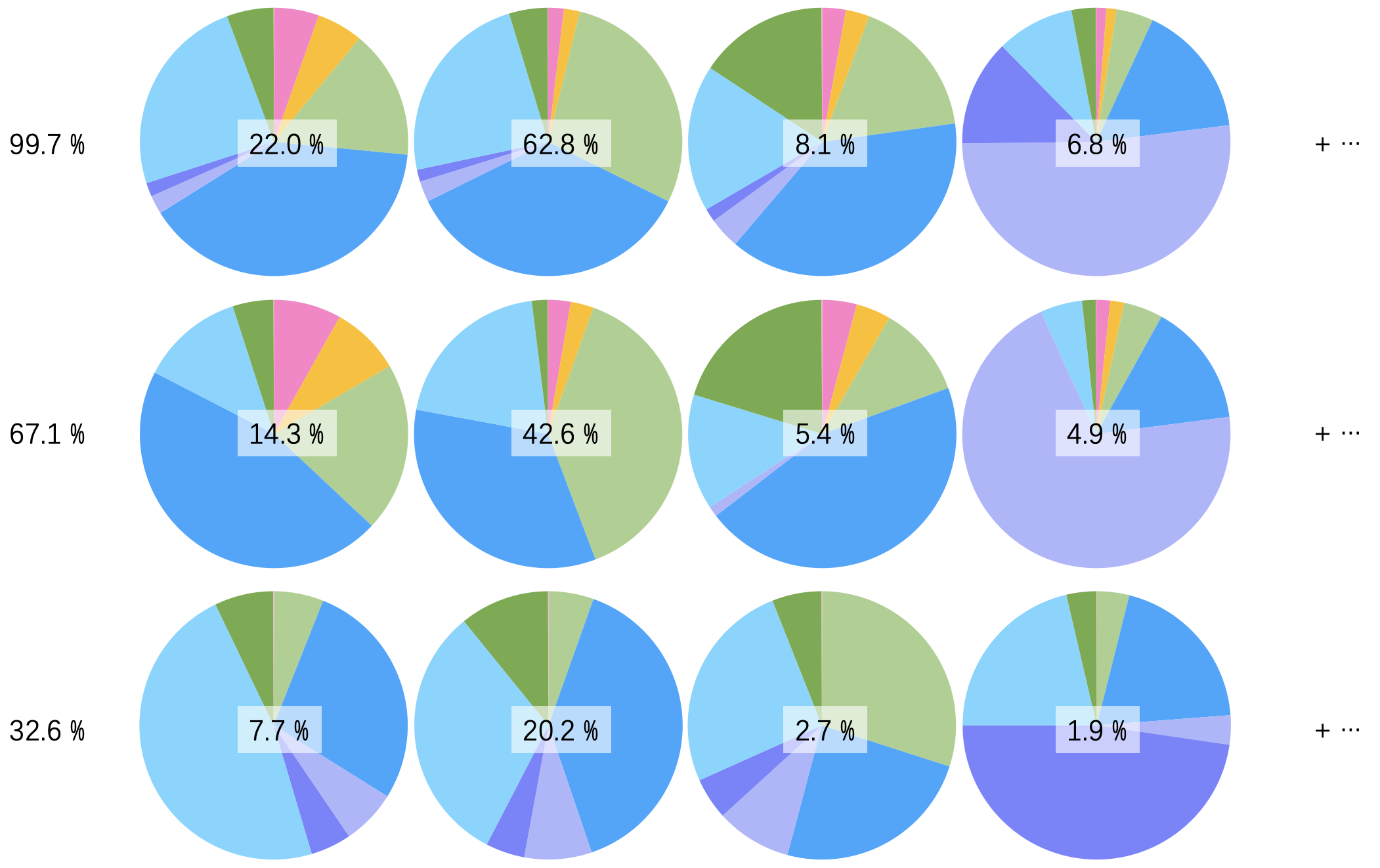}
  \includegraphics[width=0.76\textwidth]{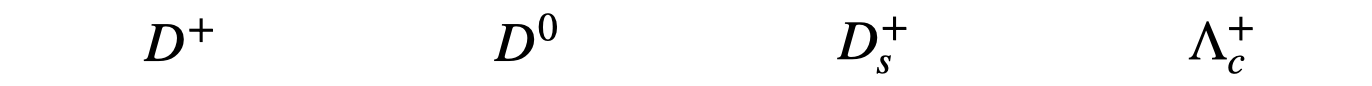}
  \includegraphics[width=\textwidth]{./figures/distributions/leg_LP.png}
  \caption{The percentage distribution pie plots of final state leading charged particles for \cjet{} that from inclusive source (top), from heavy hadron decay (middle), and from QCD fragmentation (bottom) by \Whizard. The four columns distinguish different types of leading \chadron. About 67\% of final state leading charged particles are from \chadron{} decay: \Dp, \Do, \Dsp, \Lambdacp, with proportion of 22\%/63\%/8\%/7\%, respectively. }
  \label{fig:c}
\end{figure}

\begin{figure}[htbp]
  \centering
  \includegraphics[width=\textwidth]{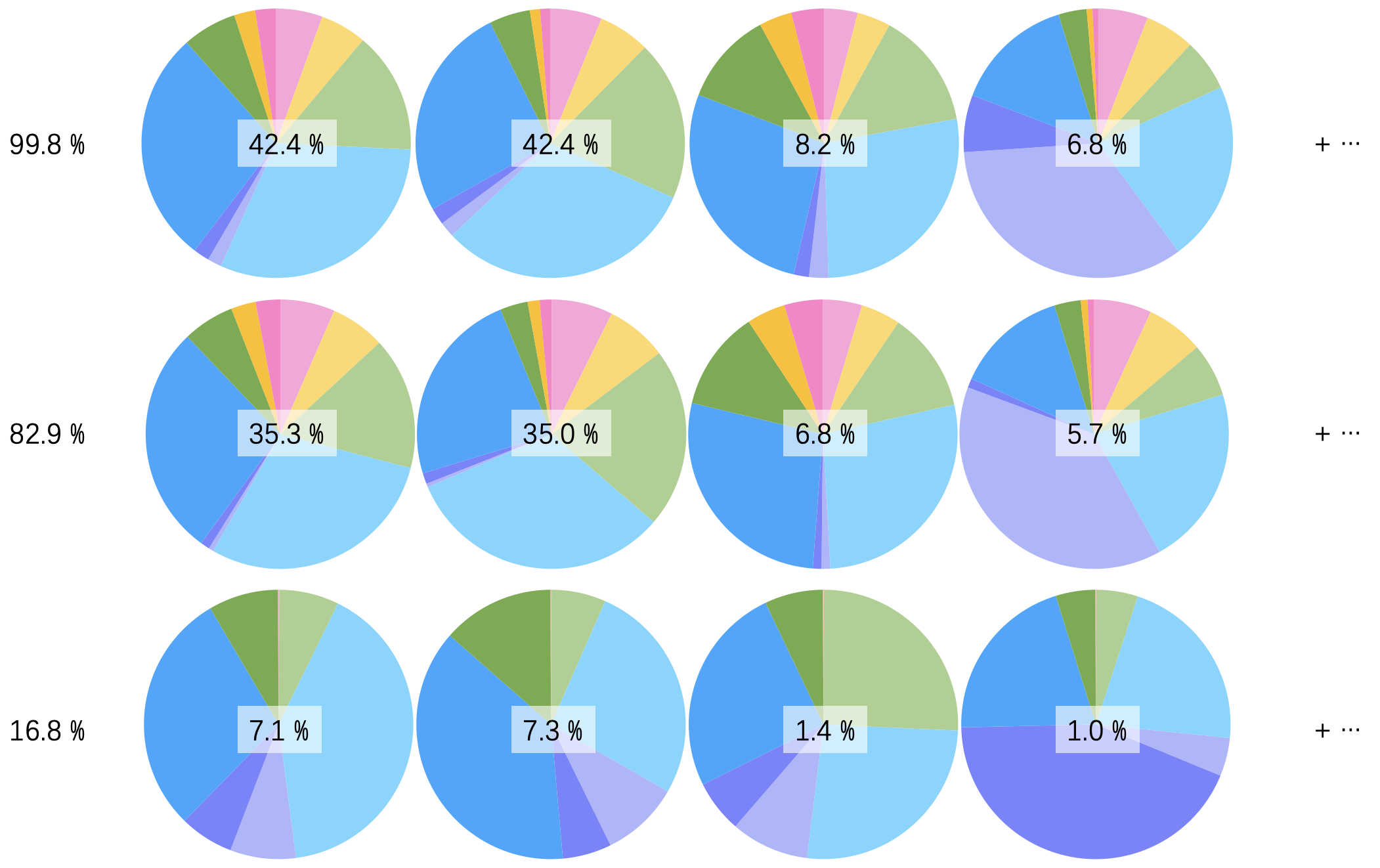}
  \includegraphics[width=0.76\textwidth]{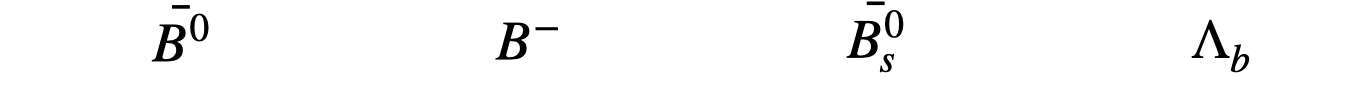}
  \includegraphics[width=\textwidth]{./figures/distributions/leg_LP.png}
  \caption{The percentage distribution pie plots of final state leading charged particles for \bjet{} that from inclusive source(top), from heavy hadron decay(middle), and from QCD fragmentation(bottom) by \Whizard. The four columns distinguish different types of leading \bhadron. About 83\% of final state leading charged particles are from \bhadron{} decay: \Bobar, \Bn, \Bsobar, \Lambdabo, with proportion of 42\%/42\%/8\%/7\%, respectively. }
  \label{fig:b}
\end{figure}

The jet charge performances for each decay source can be observed in the percentage distribution pie plot, see figure~\ref{fig:c} and~\ref{fig:b}, for the \cbjet, respectively, using the same methodology as explained in section~\ref{sec:LPJC Algorithm}.
The three rows in the figures categorize decay sources into inclusive, leading heavy hadron, or QCD fragmentation. 
The percentage distribution of final state leading charged particles from inclusive source is the sum of that from heavy hadron decay and that from QCD fragmentation.
The four columns distinguish each type of the leading heavy hadrons within the \cbjet{} into sub-groups: namely
\Dp, \Do, \Dsp, \Lambdacp{} for \cjet{} (22\%/63\%/8\%/7\% proportionally),
and \Bobar, \Bn, \Bsobar, \Lambdabo{} for \bjet{} (42\%/42\%/8\%/7\% proportionally).
The corresponding values of \etp{} for each decay source are presented in Table~\ref{tab:ETP_LPJC_OH}.
The "proportion" represents the fragmentation of each heavy hadrons.
{\color{black}The total \etp{} of 22.1\%/11.0\% for \cbjet{} is the sum of \etp{} of each sub-group weighted by the fragmentation fractions, as each leading heavy hadron type is independent of the others.}
Note that this differs from the value of 20\%/9\% for the inclusive source in Table~\ref{tab:ETP_LPJC}. 

The percentage distribution pie plots of final state leading charged particles vary with the species of leading heavy hadron.
In comparing the performance of jet charge identification for each leading heavy hadron type, 
we observe:

\begin{itemize}
\item Lepton has a probability of 5.4\% and 15.2\% to be the leading jet particle for \cbjet, respectively.
Once generated in \cjet, the charges of these leading leptons faithfully represent the jet charge since they are almost completely generated from leptonic/semi-leptonic decay of leading heavy hadrons, leading to an effective tagging power of 5.2\%, contributed evenly from muons/anti-muons and electrons/positrons.
Generated in \bjet, because the leading leptons can be generated not only from the decay of \bquark, but also the decay from \cquark{} generated in \bquark{} cascading decay, the leading lepton has a typical \mromega{} of 25.5\%; therefore the leading leptons contribute to an \etp{} of 3.6\%.
\item The leading kaons, with a probability of 28.4\% and 21.8\% to be the leading jet particle for \cbjet, achieve an \etp{} of 10.4\%/4.4\% for \cbjet.
The leading kaons decayed from \Dp and \Do{} deliver low \mromega{} of 20.4\%/4.7\%.
However, the kaons from $D_s$ can be decayed either from \cquark{} or $s$ quark and have opposite charge performances, resulting in a high \mromega{} of 35.2\%.
The \mromega{} of leading kaons from inclusive source is 19.7\%/27.5\% for \cbjet.
\item Pions have the largest probability of 37.3\% and 56.2\% to be the leading jet particle for \cbjet, 
however, its \mromega{} is the highest, equals to 38.8\%/46.3\% for \cbjet{} due to their complete sources, resulting in an \etp{} of 2.9\%/0.3\% for \cbjet.
\item With a probability of 8.9\%/6.7\%, leading protons achieve an \etp{} of 1.7\%/0.5\% for \cbjet.
Using protons from \Lambdac{} and \Lambdab{} decay yields a very small \mromega{} of 0.1\%/2.8\%, respectively, while the protons from QCD have opposite charge information.
\item Furthermore, for \bjet, the protons have opposite charge performances because the baryon number is conserved in baryons decay.
\item If the final state leading charged particle is from \Bs, the \etp{} vanishes due to its fast oscillation~\cite{LHCb:2021moh, Lenz:2006hd, CDF:2006imy, LHCb:2013lrq, ALEPH:1992net, Carrasco:2014nda}. 
However, the lepton and charged kaon with maximum momentum in the opposite direction of the \Bs, as well as the charged kaon flying in a similar direction as the \Bs, can achieve a good \mromega{} of 22.5\% and \etp{} of 20.2\%~\cite{Li:2022tlo}.
\end{itemize}

To summarize, the correlation between the final state leading charged particle and jet charge strongly depends on their origin, and the \mromega{} takes value from 0 (excellent identification) to 0.5 (no identification power). 
For final state leading charged particle of leptons inside \cjet, the \mromega{} could be reduced almost to zero, while the final state leading charged particle decay from \Bs{} is not sensitive to the jet charge at all due to the fast oscillation of \Bs. 
The \mromega{} depends strongly on whether they are generated from leading heavy hadron decay or QCD fragmentation.
The information regarding the source can be extracted from a well-defined vertex (VTX) due to the long decay length of leading heavy hadron, see Table~\ref{tab:ETP_LPJC_OH}.
If the VTX can effectively differentiate between protons originating from heavy hadron decay and those from QCD fragmentation, LPJC can achieve
{\color{black}an upper limits of}
\etp{} of 35.7\%/11.2\% for \cbjet.

\section{Weighted Jet Charge method (WJC)}
\label{sec:WJC}

The Weighted Jet Charge method (WJC) is based on the calculation of jet charge as described in ref.~\cite{Waalewijn:2012sv}.
In this research, the jet charge $Q^\kappa_\mathrm{jet}$ is calculated by the energy-weighted sum of the electric charges of the particles within a jet,
which is defined as
\begin{equation}
\label{equ:weight}
    Q^\kappa_\mathrm{jet} = \frac{\sum_i E_i^\kappa \ Q_i}{\sum_i E_i^\kappa},
\end{equation}
where the sum runs over all charged particles in a hemisphere to a particular jet.
The parameter $\kappa$ is responsible for managing the relative contribution of both low and high-energy particles to the jet charge.

\begin{figure}[htbp]
  \centering
  \includegraphics[width=0.4\textwidth]{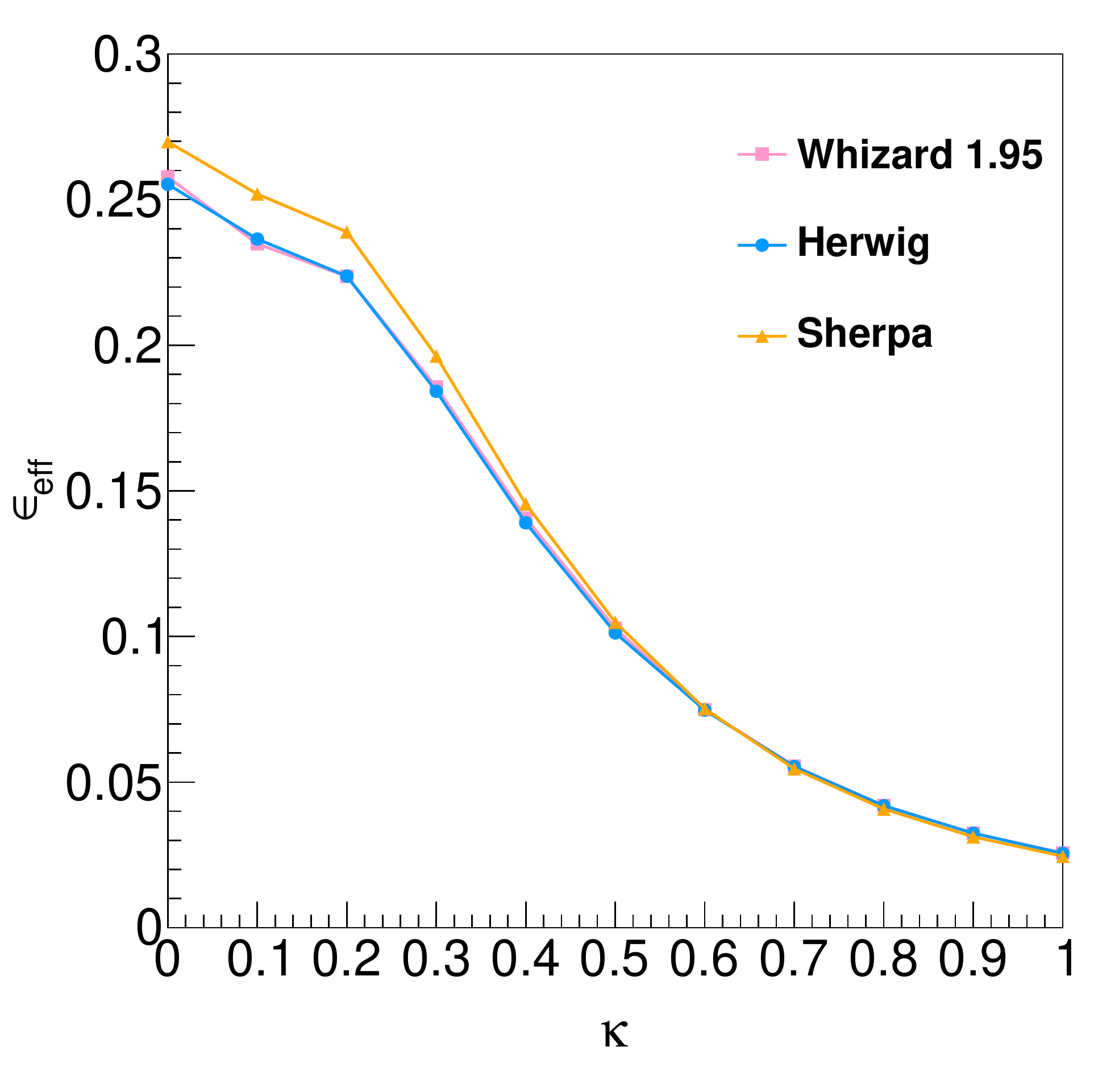}
  \hspace{0.8cm}
  \includegraphics[width=0.4\textwidth]{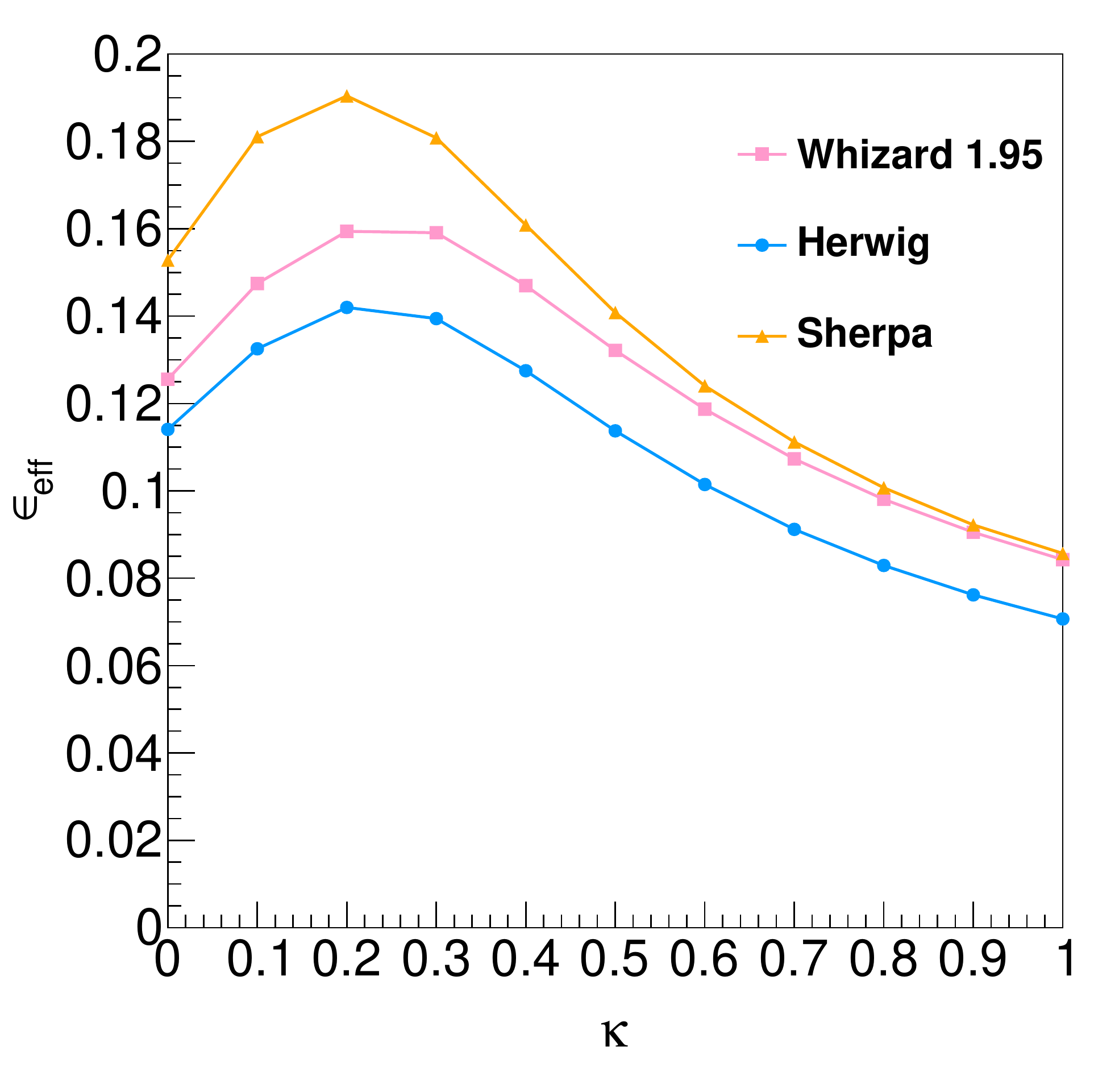}
  \caption{{\color{black}The WJC \etp{} with different $\kappa$ (from 0 to 1.0) for \cjet{} (left) and \bjet{} (right). The optimal $\kappa$ is 0 for \cjet{} and 0.2 for \bjet.}}
  \label{fig:ETP_WJC_vs_Generator}
\end{figure}

\begin{figure}[htbp]
  \centering
  \includegraphics[width=0.4\textwidth]{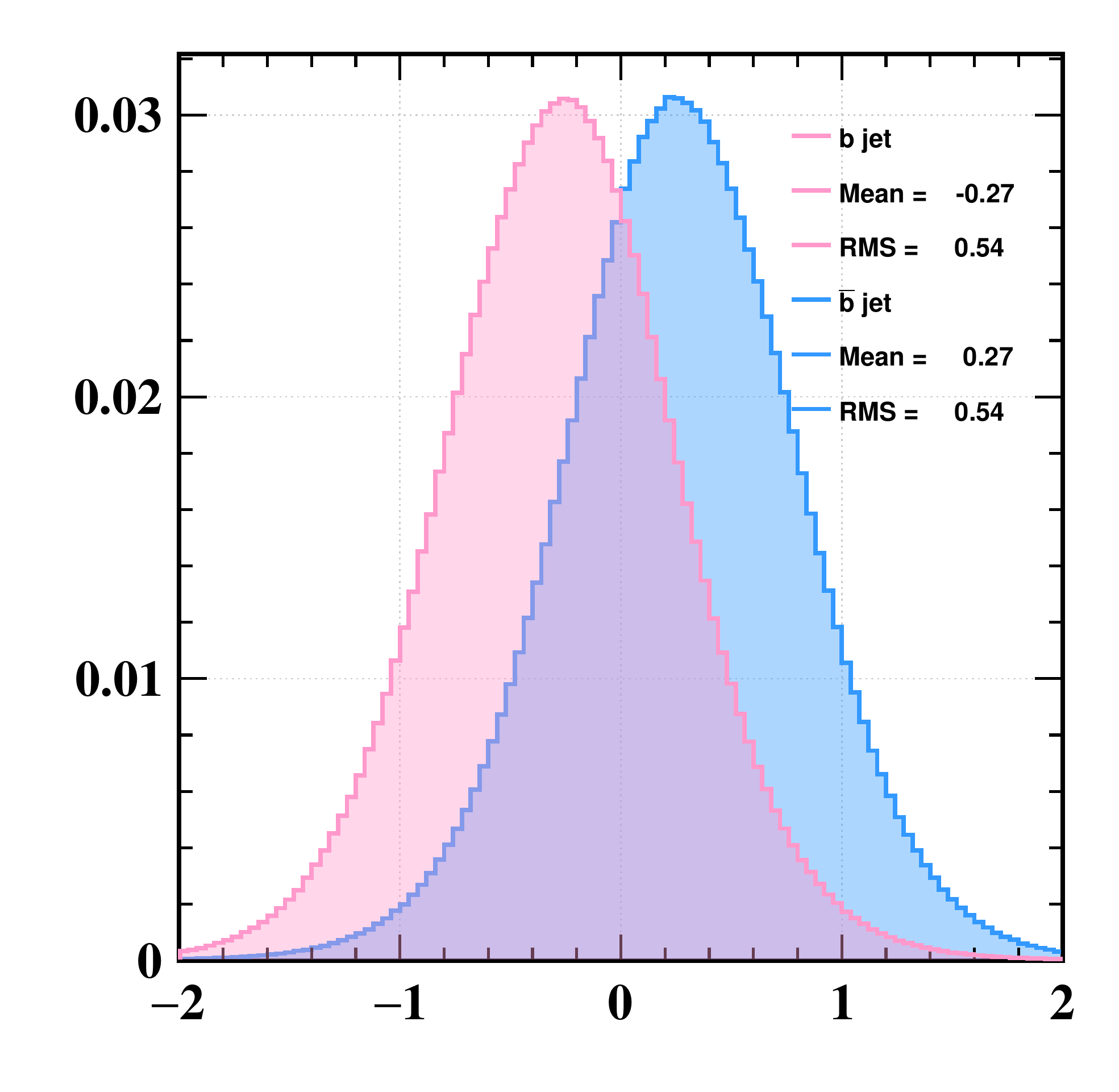}
  \caption{The charge distributions for \bjet{} with $\kappa = 0.2$.}
  \label{fig:charge_kappa02}
\end{figure}

Using the conventional WJC, the variation of \etp{} with different $\kappa$ values is illustrated in figure~\ref{fig:ETP_WJC_vs_Generator}.
The three convex lines by three generators exhibit a consistent trend.
{\color{black}The differences in generators may be due to the different non-perturbative hadronization models used to explain the transition from partonic states to the final hadronic states: phenomenological "string models" (\Pythia) or "cluster models" (\Herwig and \Sherpa) are used for calculations, and the different secondary decay models.
This might lead to different proportion of final state charged particles, thus affecting the \etp.
The real fragmentation will be determined with real data in future.}
The optimal $\kappa$ values with \Whizard{} are found to be 0/0.2 for \cbjet, yielding \etp{} values of 25.8\%/15.9\%, respectively, as detailed in Table~\ref{tab:ETP_WJC}. 
The jet charge distributions, using \bjet{} and \bbarjet{} as examples, are presented in figure~\ref{fig:charge_kappa02}. 

\begin{table}[htb]
    \centering
    \renewcommand{\arraystretch}{1.2}
	\caption{The WJC \etp{} for the \cbjet{} is determined using three generators, where $\kappa$ is respectively set to the optimal value.}
    \begin{tabular}[t]{|c|c|c|c|}
        \hline
         & \Whizard & \Herwig & \Sherpa \\
        \hline
        \cjet  & $25.8\%$ & $25.5\%$ & $27.0\%$ \\
        \hline
        \bjet & $15.9\%$ & $14.2\%$ & $19.0\%$ \\
        \hline
    \end{tabular}
    \label{tab:ETP_WJC}
\end{table}

\section{Combination}
\label{sec:Combination}

\subsection{Improved Weighted Jet Charge method (IWJC)}
\label{sec:IWJC}


The optimal $\kappa$ value differs for each type of final state leading charged particles.
The difference in $\kappa$ values for \bcjet s is due to the different masses and decay properties.
{\color{black}Specifically, the parameter $\kappa$ determines the relative contribution of low- and high- energy particles to the jet charge.
In other words, it quantifies how the jet charge is influenced by the energy of the individual particles within the jet.}
The \bquark{} has a longer lifetime compared to \cquark, leading to a different energy response compared to \cjet s.
The same is true for different heavy hadrons in {\color{black}Table~\ref{tab:kappa_FP}}.
Different decay channels of \bcjet s may have distinct energy distributions, necessitating different $\kappa$ values to accurately capture the charge distribution within the jet.
Our analysis has accounted for these differences in $\kappa$ values by calibrating the jet energy response separately for each case. 
This ensures that the measured jet properties are effectively attributed to the corresponding quark flavor.

Therefore, distinguishing different types of final state leading charged particles can improve the jet charge performance.
In addition, when $\kappa = 0$, compared to other $\kappa$ values, the \mromega{} might be smaller as a cost of efficiency.
In this case, the samples could be further categorized into two groups:
whether $Q^{\kappa=0}_\mathrm{jet}=0$ or not.
By optimizing $\kappa$ for each type of final state leading charged particles 
and re-optimize $\kappa$ for those $Q^{\kappa=0}_\mathrm{jet}=0$, 
we developed an Improved Weighted Jet Charge method (IWJC).

Jets can be categorized into several classes based on the species and origin of their leading charged particles. 
The optimal $\kappa$ for each category can differ significantly from that of others. 
Therefore, calculating the jet charge using $\kappa$ values determined from the species and origin of these leading particles can enhance identification performance. 
Additionally, in cases where $\kappa$ is zero, the jet charge may coincidentally become zero. 
In such instances, a re-optimized $\kappa$ is utilized. 
To incorporate these improvements, we have developed an Improved Weighted Jet Charge method (IWJC).
The optimal parameter $\kappa$ varies with the species and origin, summarized in Table~\ref{tab:kappa_FP}.

\begin{table}[htb]
    \centering
    \renewcommand{\arraystretch}{1.2}
	\caption{The optimal parameter $\kappa$ for each type of final state particles from different sources. We consider the $\kappa$=100 as $+\infty$, which corresponds to LPJC method.}
    \begin{tabular}[t]{|c|c|c|c|c|}
        \hline
        & & Inclusive & Heavy hadron & QCD \\
        \hline
        \multirow{6}{*}{\cjet} & $e$ & $+\infty$ & $+\infty$ & - \\
        \cline{2-5}
        & $\mu$ & $+\infty$ & $+\infty$ & - \\
        \cline{2-5}
        & $K$ & $+\infty$ & $+\infty$ & 0.3 \\
        \cline{2-5}
        & $\pi$ & 0 & 0 & 0 \\
        \cline{2-5}
        & $p$ & 0.3 & $+\infty$ & 0 \\
        \hline
        \multirow{6}{*}{\bjet} & $e$ & 0.4 & 0 & - \\
        \cline{2-5}
        & $\mu$ & 0.4 & 0 & - \\
        \cline{2-5}
        & $K$ & 0.4 & 0 & $+\infty$ \\
        \cline{2-5}
        & $\pi$ & 0.2 & 0 & 0 \\
        \cline{2-5}
        & $p$ & 0 & $+\infty$ & 0.1 \\
        \hline
    \end{tabular} 
    \label{tab:kappa_FP}
\end{table}

For $e, \mu, K, \pi$ from \bhadron{} decay, as well as $\pi$ from \chadron{} decay, the optimal $\kappa$ approaches 0, corresponding to the sum of jet particle charge.
On the other hand, for protons from \bhadron{} decay and $e, \mu, K, p$ from \chadron{} decay, the optimal $\kappa$ approaches $+\infty$, which means the reconstructed jet charge will be determined by the jet particle with highest energy - corresponding to LPJC method.
Using IWJC, we determined a benchmark value of the \etp{} in Table~\ref{tab:ETP_IWJC}.

\begin{table}[htb]
    \centering
    \renewcommand{\arraystretch}{1.2}
	\caption{The IWJC \etp{} at the \cbjet{} using three generators.}
    \begin{tabular}[t]{|c|c|c|c|}
        \hline
         & \Whizard & \Herwig & \Sherpa \\
        \hline
        \cjet  & $31.1\%$ & $31.7\%$ & $33.1\%$ \\
        \hline
        \bjet & $18.3\%$ & $16.3\%$ & $20.0\%$ \\
        \hline
    \end{tabular}
    \label{tab:ETP_IWJC}
\end{table}

\subsection{Heavy Flavor Jet Charge method (HFJC)}
\label{sec:HFJC}

Although the inclusive performance of LPJC is inferior to that of IWJC, LPJC does have advantages in specific categories of final state leading charged particle species. 
Therefore, an appropriate combination of LPJC and IWJC could lead to improved jet charge identification performance compared to either of them individually. 
To this end, we have developed a combined jet charge algorithm, Heavy Flavor Jet Charge method (HFJC), which can be divided into six steps:

\begin{enumerate}
\item Categorize final state leading charged particles into five types: \emuKpipr.
\item Perform LPJC and IWJC identification in each sub-category. 
\item Categorize each type of final state leading charged particles into three groups: two decisions agree, two decisions disagree, and only one method has a decision.
\item For each group, calculate the corresponding \mromega.
\item Calculate the \etp{} of each group using eq.~\eqref{equ:combine}.
\item Add the \etp{} of each type of final state particles, which form an almost full coverage of all the events, to obtain the total \etp.
\end{enumerate}  

According to refs. \cite{Heinicke:2229990, Belle-II:2018jsg}, the combined \etp{} can be calculated by
\begin{equation}
\label{equ:combine}
\epsilon_{\mathrm{eff}}^{\mathrm{comb}} = 
\sum_{i=1}^{\mathrm{N_{method}}} s_{i} \ \xi_{i} \ (1-2\omega_{i})^2,
\end{equation}
where 
$s_{i}$ is the decision weight of $i$-th method.
$\omega_{i}$ is the \mromega{} of $i$-th method.
$\xi_{i}$ is the tagging decision of $i$-th method.
The variable $\xi_i$ represents the decision of a particular judgment, which can take values of 
+1 when the input is tagged as a positive jet,
-1 when the input is tagged as a negative jet,
0 when the input is untagged. 

The HFJC \etp{} values of each type of final state leading charged particles are shown in Table~\ref{tab:ETP_HFJC}.
Compared to LPJC and WJC, 
the leading leptons for \cjet{} still deliver a \mromega{} of close to zero.
The \etp{} of leading pions, kaons, and protons improved tens or hundreds of times.
The total \etp{} improved about 90\%/130\% for \cbjet.

The \etp{} of each method is summarized in Table~\ref{tab:ETP_all_methods}.
Each column compares each method: Leading Particle Jet Charge (LPJC), conventional Weighted Jet Charge (WJC), Improved Weighted Jet Charge method (IWJC), Heavy Flavor Jet Charge (HFJC), and ideal cases that the origin of final state leading charged particle can be distinguished correctly, and that \bchadron{} type can be reconstructed correctly.
Each row compares the three generators: \Whizard, \Herwig, and \Sherpa.
The tendencies between three generators of the improvement of jet charge methods are almost consistent,
and the differences between the three generators of the same method are below 11\%/12\% for \cbjet{} relative to \Whizard.

At \bjet, using the energy-weighted sum of jet charge, WJC exhibits better jet charge performance than LPJC.
While for \cjet, for final state \emuK, LPJC exhibits better jet charge performance.
IWJC categorizes final state particles to improve WJC, reaching an improvement of 15\%/20\%.
By combing the methods above, HFJC achieves an \etp{} of 39.0\%/20.4\% for \cbjet.
If the origin of the final state leading charged particle can be distinguished between leading heavy hadron decay and QCD fragmentation, the \etp{} could improve to 45\%/37\% for \cbjet, reaching an improvement of 15\%/80\%.
If each type of leading heavy hadrons can be distinguished, the \etp{} reaches 56.1\%/47.8\% for \cbjet.

\begin{table}[htb]
    \centering
    \renewcommand{\arraystretch}{1.2}
	\caption{The HFJC \etp{} of each type of final state leading charged particles and the total \etp{} at the \cbjet.}
    \begin{tabular}[t]{|c|c|c|c|c|c|c|}
        \hline
        & & \multicolumn{3}{c|}{\Whizard} & \Herwig & \Sherpa \\
        \hline
        & & {\color{black}$\epsilon_{\mathrm{tag,max}}$} & $\omega$ & $\epsilon_{\mathrm{eff}}$ & $\epsilon_{\mathrm{eff}}$ & $\epsilon_{\mathrm{eff}}$ \\
        \hline
        \multirow{6}{*}{\cjet} & $e$ & $2.7\%$ & $1.9\%$ & $2.5\%$ & $3.4\%$ & $2.9\%$ \\
        \cline{2-7}
        & $\mu$ & $2.7\%$ & $0.5\%$ & $2.7\%$ & $3.0\%$ & $3.2\%$ \\
        \cline{2-7}
        & $K$ & $28.4\%$ & $15.2\%$ & $13.8\%$ & $14.5\%$ & $14.7\%$ \\
        \cline{2-7}
        & $\pi$ & $57.3\%$ & $22.9\%$ & $16.8\%$ & $17.2\%$ & $17.7\%$ \\
        \cline{2-7}
        & $p$ & $8.9\%$ & $19.8\%$ & $3.2\%$ & $2.2\%$ & $2.3\%$ \\
        \cline{2-7}
        & Sum & $100\%$ & $18.8\%$ & $39.0\%$ & $18.3\%$ & $40.3\%$ \\
        \hline
        \multirow{6}{*}{\bjet} & $e$ & $7.6\%$ & $21.6\%$ & $2.5\%$ & $2.4\%$ & $2.4\%$ \\
        \cline{2-7}
        & $\mu$ & $7.6\%$ & $21.2\%$ & $2.5\%$ & $2.5\%$ & $2.4\%$ \\
        \cline{2-7}
        & $K$ & $21.8\%$ & $23.8\%$ & $6.0\%$ & $5.1\%$ & $6.2\%$ \\
        \cline{2-7}
        & $\pi$ & $56.2\%$ & $31.1\%$ & $8.1\%$ & $7.1\%$ & $10.2\%$ \\
        \cline{2-7}
        & $p$ & $6.7\%$ & $28.1\%$ & $1.3\%$ & $2.1\%$ & $1.3\%$ \\
        \cline{2-7}
        & Sum & $100\%$ & $27.4\%$ & $20.4\%$ & $19.1\%$ & $22.5\%$ \\
        \hline
    \end{tabular} 
    \label{tab:ETP_HFJC}
\end{table}

\begin{table}[htb]
    \centering
    \renewcommand{\arraystretch}{1.2}
	\caption{The \etp{} at the \cbjet{} of each method: Leading Particle Jet Charge (LPJC), conventional Weighted Jet Charge (WJC), Improved Weighted Jet Charge method (IWJC), Heavy Flavor Jet Charge (HFJC), and ideal cases that the origin of final state leading charged particle can be distinguished correctly, and that \bchadron{} type can be reconstructed correctly. Results are presented using three generators for comparison: \Whizard, \Herwig, and \Sherpa.
}
    \resizebox{\textwidth}{!}{
    \begin{tabular}[t]{|c|c|c|c|c|c|c|c|}
        \hline
        $\epsilon_{\mathrm{eff}}$ & Generator & LPJC & WJC & IWJC & HFJC & \makecell{HFJC$^{\ast}$ \\ (distinguish \\ \bchadron \\ /QCD)} & \makecell{HFJC$^{\ast\ast}$ \\ (distinguish \\ \bchadron \\ /QCD \\ and \\ \bchadron{} type)} \\
        \hline
        \multirow{3}{*}{\cjet} & \Whizard & 20.2\% & 25.8\% & 31.1\% & 39.0\% & 45.0\% & 56.1\% \\
        \cline{2-8}
        & \Herwig & 22.5\% & 25.5\% & 31.7\% & 40.3\% & 46.8\% & 57.3\% \\
        \cline{2-8}
        & \Sherpa & 21.5\% & 27.0\% & 33.1\% & 40.8\% & 48.6\% & 59.4\% \\
        \hline
        \multirow{3}{*}{\bjet} & \Whizard & 8.9\% & 15.9\% & 18.3\% & 20.4\% & 36.9\% & 47.8\% \\
        \cline{2-8}
        & \Herwig & 8.4\% & 14.2\% & 16.3\% & 19.1\% & 36.0\% & 46.6\% \\
        \cline{2-8}
        & \Sherpa & 7.8\% & 19.0\% & 20.0\% & 22.5\% & 35.8\% & 48.9\% \\
        \hline
    \end{tabular}
    }
    \label{tab:ETP_all_methods}
\end{table}

\section{Impact of realistic detector effects}
\label{sec:Detector}

{\color{black}The jet charge performances described in the previous sessions are obtained using truth level information, which, to some extent, corresponds to an ideal detector that has 100\% acceptance, 100\% reconstruction efficiency for all final state charged particles, ideal flavor tagging, PID identification and momentum measurements. 
This section explores the disparity between the actual \etp{} and the maximum attainable limit.}

The jet charge performance will necessarily depend on the specific detector configuration and performance in actual experiments at the high energy frontier.
It is necessary to understand and quantify the influence of the detector parameters.
We analyze the influence on high energy frontier and Higgs factory detectors in references~\cite{EuropeanStrategyforParticlePhysicsPreparatoryGroup:2019qin,ECFA2021,Gaudio:2022jve}, and typically on CEPC baseline detector~\cite{CEPCStudyGroup:2018ghi}.
The relevant detector effects  are mainly as follows.

\begin{itemize}
\item \textbf {Jet Flavor Tagging (FT)} 

The jet charge performances described above are achieved at pure samples; 
In realistic HEP experiments, it is usually difficult to distinguish jets initiated by different species of quarks or gluon with high efficiency/purity. 

The performance of FT can be characterized by a parameter matrix, which converts the truth level (T) to reconstructed level (re) for heavy flavor jet, defined as

\begin{equation}
\begin{pmatrix}
    b^{re} \\
    c^{re} \\
\end{pmatrix}
=
\begin{pmatrix}
    1-\Omega_{cb} & \Omega_{cb} \\
    \Omega_{bc} & 1-\Omega_{bc}
\end{pmatrix}
\begin{pmatrix}
    b^{T} \\
    c^{T} \\
\end{pmatrix},
\end{equation}

where $\Omega$ is the misjudgment rate of flavor tagging.
For simplicity, we consider the FT matrix as a single-parameter matrix, 
\begin{equation}
\begin{pmatrix} 
    1-\Omega & \Omega \\
    \Omega & 1-\Omega
\end{pmatrix}.
\end{equation}

The efficiency of \bcjet{} can be expressed as 1-$\Omega$.


Combing with quark charge information, the performance of the flavor-charge tagging (FC) can be characterized by a parameter matrix, defined as

$$
\begin{pmatrix}
    b^{re} \\
    \bar{b}^{re} \\
    c^{re} \\
    \bar{c}^{re}
\end{pmatrix}
=
\begin{pmatrix}
    r_{bb} & r_{\bar{b}b} & r_{cb} & r_{\bar{c}b} \\
    r_{b\bar{b}} & r_{\bar{b}\bar{b}} & r_{c\bar{b}} & r_{\bar{c}\bar{b}} \\
    r_{bc} & r_{\bar{b}c} & r_{cc} & r_{\bar{c}c} \\
    r_{b\bar{c}} & r_{\bar{b}\bar{c}} & r_{c\bar{c}} & r_{\bar{c}\bar{c}}
\end{pmatrix}
\begin{pmatrix}
    b^{T} \\
    \bar{b}^{T} \\
    c^{T} \\
    \bar{c}^{T}
\end{pmatrix}.
$$

The elements in FC matrix can be calculated from flavor tagging misjudgment rate $\Omega$ and jet charge misjudgment rate $\omega$ by

\begin{center} 
\resizebox{0.8\textwidth}{!}{ 
$\begin{pmatrix}     
(1-\Omega)(1-\omega_{b})/2 & (1-\Omega)\omega_{b}/2 & \Omega \ \omega'_{c}/2 & \Omega(1-\omega'_{c})/2 \\ (1-\Omega)\omega_{b}/2 & (1-\Omega)(1-\omega_{b})/2 & \Omega(1-\omega'_{c})/2 & \Omega \ \omega'_{c}/2 \\ \Omega \ \omega'_{b}/2 & \Omega(1-\omega'_{b})/2 & (1-\Omega)(1-\omega_{c})/2 & (1-\Omega)\omega_{c}/2 \\ \Omega(1-\omega'_{b})/2 & \Omega \ \omega'_{b}/2 & (1-\Omega)\omega_{c}/2 & (1-\Omega)(1-\omega_{c})/2
\end{pmatrix}$ .
}
\end{center}

where, the $\omega'_{b(c)}$ represents the jet charge misjudgment rate for $b(c)$jet that we reconstructed as $c(b)$jet.
Through the charge correlation between final state particles and \bcquark{} presented in section ~\ref{sec:LPJC}, for $e, \mu, \pi$, $\omega'_{b(c)} = 1-\omega_{b(c)}$;
for $K,p$, $\omega'_{b(c)} = \omega_{b(c)}$.

The misjudgment rate of jet charge with flavor mixing, $\omega^{\mathrm{FC}}$, can be obtained from the FC matrix
\begin{equation}
\begin{split}
\omega^{\mathrm{FC}}_{b} = (1-\Omega) \omega_{b} + \Omega(1-\omega'_{c}) \,,
\qquad
\omega^{\mathrm{FC}}_{c} = (1-\Omega) \omega_{c} + \Omega(1-\omega'_{b}) \,.
\end{split}
\end{equation}


The corresponding effective tagging power $\epsilon_{\mathrm{eff}}^{\mathrm{FC}}$ can be expressed as
\begin{equation}
\begin{split}
\epsilon_{\mathrm{eff},b}^{\mathrm{FC}} = (1-\Omega) (1-2\omega^{\mathrm{FC}}_{b})^2 \,,
\qquad
\epsilon_{\mathrm{eff},c}^{\mathrm{FC}} = (1-\Omega) (1-2\omega^{\mathrm{FC}}_{c})^2 \,.
\end{split}
\end{equation}

\begin{figure}[htbp]
  \centering
  \includegraphics[width=0.8\textwidth]{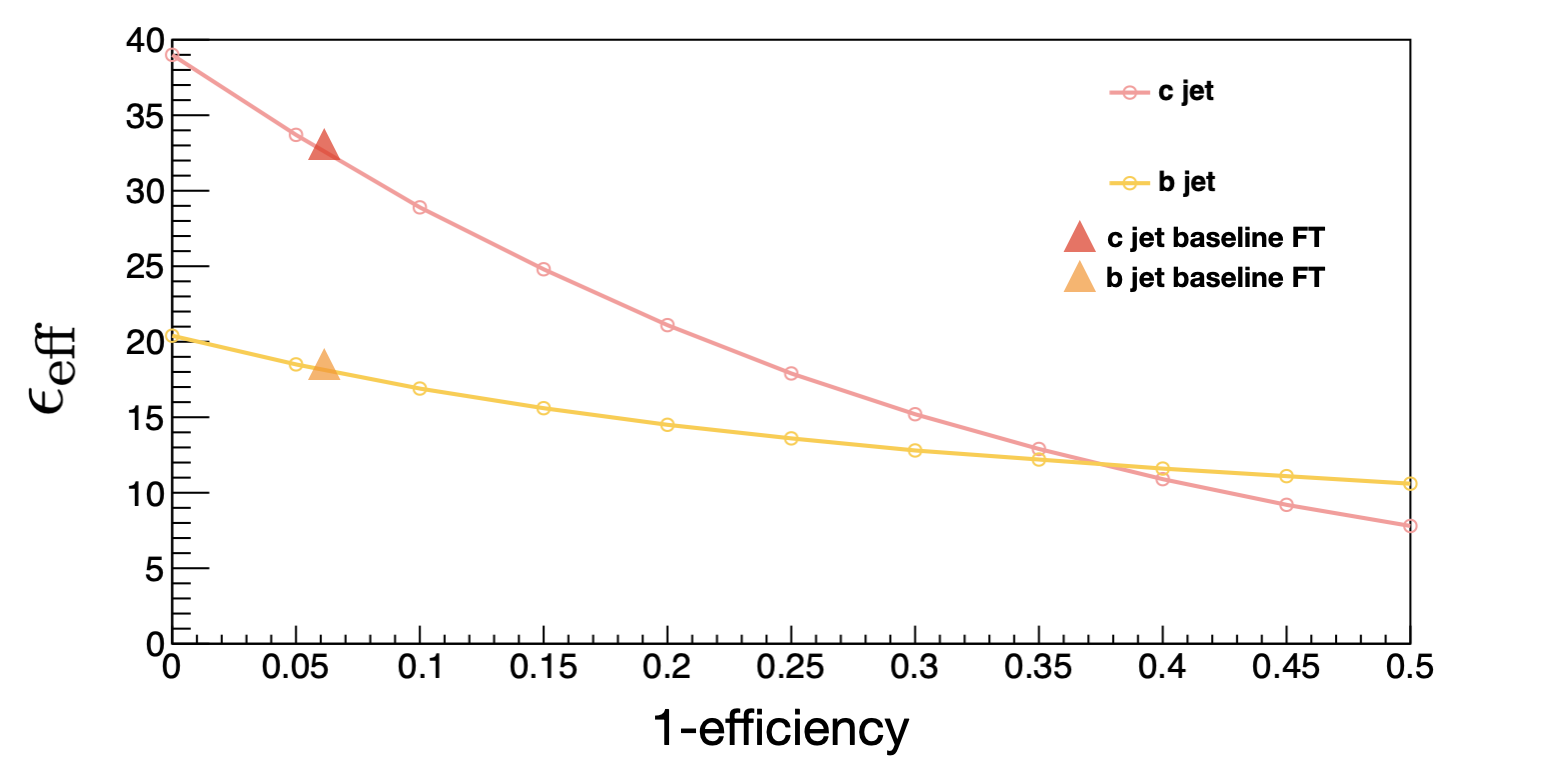}
  \caption{The \etp{} as a function of the \bcquark{} flavor tagging efficiency, using HFJC method.}
  \label{fig:ETP_vs_FT}
\end{figure}

The correlation between $\epsilon_{\mathrm{eff}}^{\mathrm{FC}}$ and FT efficiency is illustrated in figure~\ref{fig:ETP_vs_FT}, which shows a clear correlation between the flavor tagging performance and jet charge performance. 
{\color{black}At the CEPC, simulation studies show that at inclusive hadronic events at \Z{} pole operation (\Zqq), 
the \cbjet{} can be typically identified with an efficiency of
78.4\%/91.1\%~\cite{Zhu:2023xpk}, reaching an \etp{} equals to 33.8\%/18.7\% for \cbjet, respectively.
}

\item \textbf {Vertex resolution} 
The vertex information plays a crucial role in distinguishing the origin of the final state leading charged particles, namely whether they arise from heavy hadron decay or QCD fragmentation. 
The HFJC* presented in Table~\ref{tab:ETP_all_methods} is based on ideal vertex resolution. 
However, the interdependence between vertex resolution and jet charge performance is of significant importance. 
The reconstructed impact point (IP) can be used to measure vertex resolution, which represents the resolution of the decay sources of final state charged particles.
We investigate the influence of IP resolution on the HFJC* performance.

\begin{figure}[htbp]
  \centering
  \includegraphics[width=0.8\textwidth]{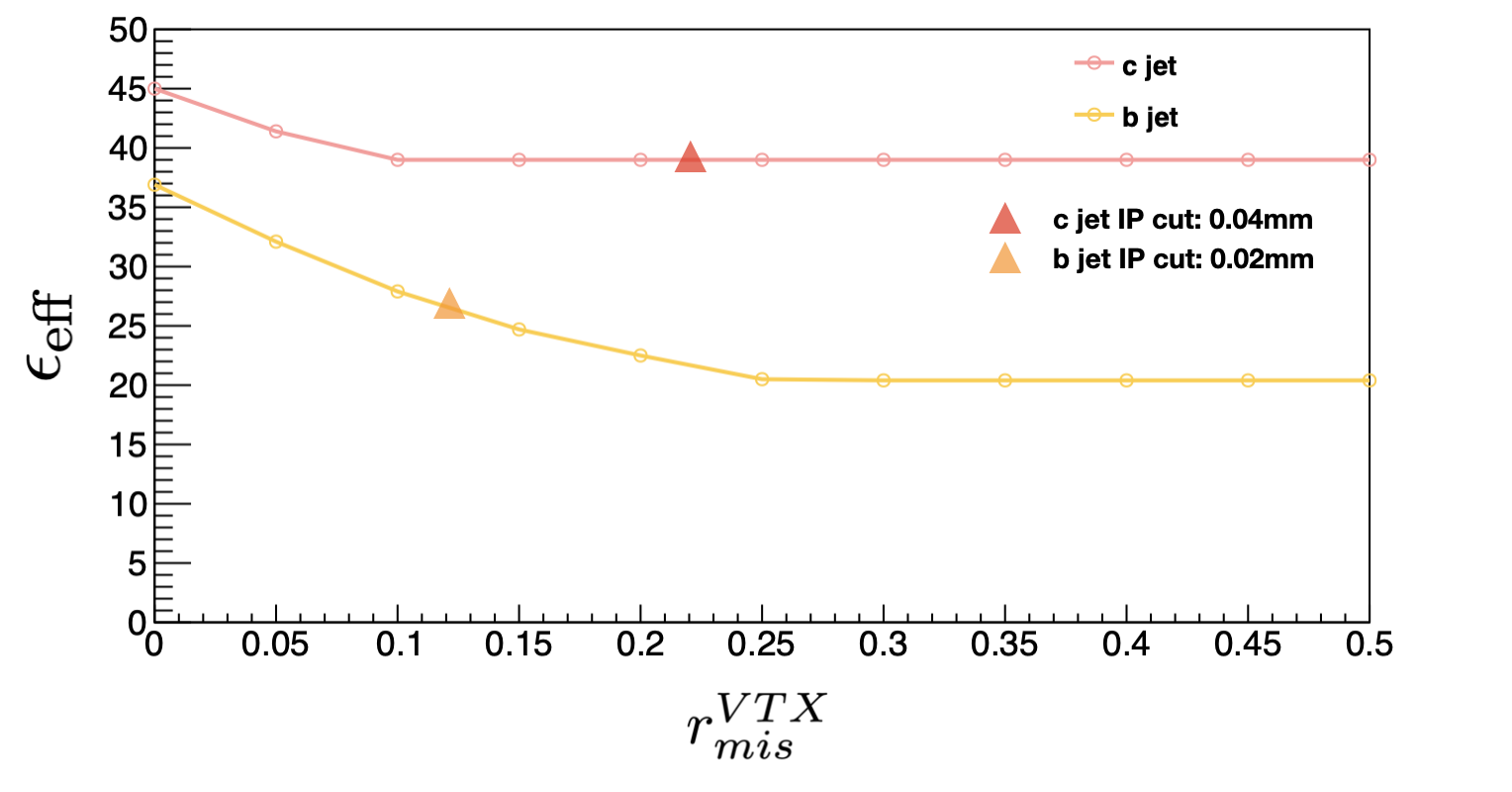}
  \caption{The HFJC* \etp{} varies with the mis-distinction rate between heavy hadron decay and QCD fragmentation ${\color{black}r_{mis}^{vtx}}$ for \bjet{} and \cjet.}
  \label{fig:ETP_vs_vtx}
\end{figure}

\begin{figure}[htbp]
  \centering
  \includegraphics[width=0.8\textwidth]{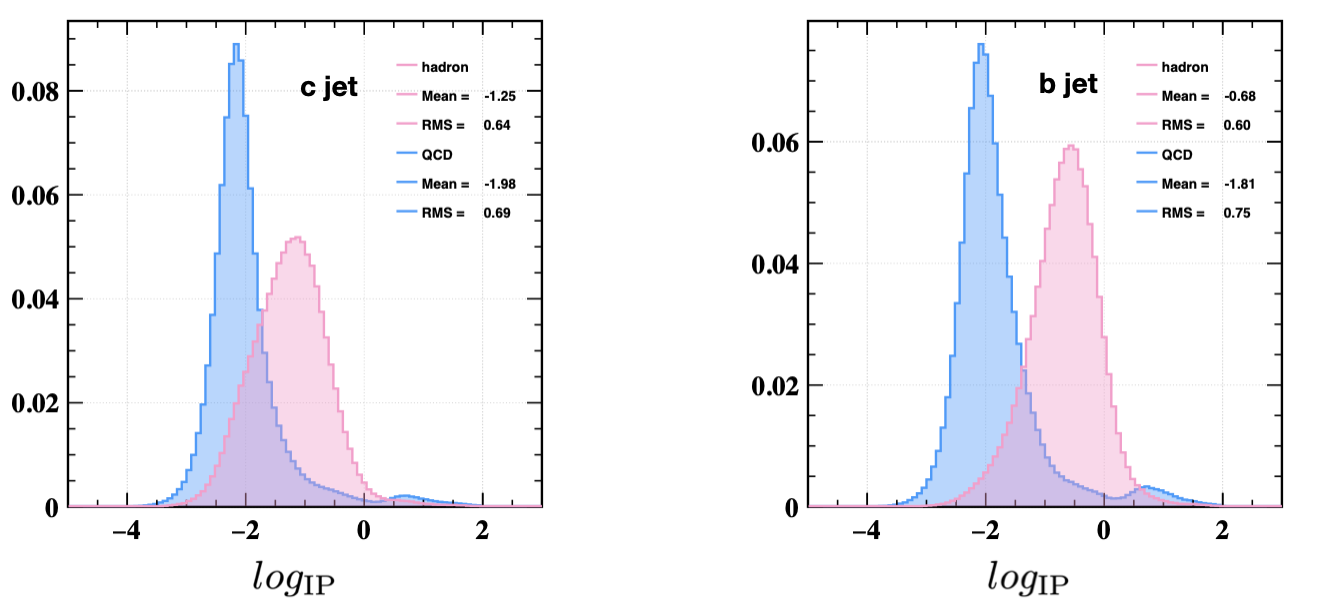}
  \caption{The reconstructed $log_{\mathrm{IP}}$ ({\color{black}the logarithm of impact parameter}) distributions of final state leading charged particles that from heavy hadron decay and QCD fragmentation for \cjet{} (left) and \bjet{} (right).}
  \label{fig:IP}
\end{figure}

For simplicity, the distinction between heavy hadron decay and QCD fragmentation using IP resolution can also be characterized by a single-parameter matrix given by
\begin{equation}
\begin{pmatrix} 
    r_T & {\color{black}r_{mis}^{vtx}} \\
    {\color{black}r_{mis}^{vtx}} & r_T 
\end{pmatrix},
\end{equation}
where $r_{T} = 1-{\color{black}r_{mis}^{vtx}}$.
The correlation between $\epsilon_{\mathrm{eff}}$ and ${\color{black}r_{mis}^{vtx}}$ is shown in figure~\ref{fig:ETP_vs_vtx}.

The \etp{} of \cbjet{} decreases rapidly due to the fact that the heavier hadron mass and larger decay length lead to a significant impact of IP resolution on the jet charge measurement. 
The IP distributions of final state leading charged particles that arise from heavy hadron decay and QCD fragmentation for \cbjet{} are displayed in figure~\ref{fig:IP}, which is scaled to 1 to correspond to the matrix element.
The peaks of higher values on these two graphs are mainly from $K_s/\Lambda$ decay.

The reconstructed IP distinctions between final state charged particles that arise from heavy hadron decay and QCD fragmentation are relatively better for \bjet{} than for \cjet, and the IP resolution has a bigger impact on \bjet{} than on \cjet{} charge identification performance.
The benchmark IP cuts of $0.02/0.04mm$ for \cbjet{} are marked with green dashed lines, which correspond to 
the triangle markers in figure~\ref{fig:ETP_vs_vtx}, leading to an \etp{} of 39.0\%/26.8\% for \cbjet.
Therefore, improving IP resolution is crucial for enhancing the performance of jet charge measurements for \bjet, and relatively less crucial for \cjet.
Further research in developing more advanced IP reconstruction techniques is worth pursuing.

\item \textbf {PID} 

Jet charge performance is sensitive to PID performance, which encompasses lepton identification and hadron identification.
\subitem \textbf {Lepton identification} 
The lepton identification reaches an efficiency and mis-identification rate of 98\% and 1\% for energy higher than 2 \GeV ~\cite{Yu:2021pxc}.
The impact of lepton identification on \etp{} can be ignored.
\subitem \textbf {Hadron identification} 
The hadron identification ($K,\pi,p$) is comparatively inferior to lepton identification; therefore, its influence on jet charge \etp{} merits investigation.
For simplicity, the hadron identification performance can also be characterized by a single-parameter matrix,
\begin{equation}
\begin{pmatrix} 
    r_T & {\color{black}r_{mis}^{pid}} & {\color{black}r_{mis}^{pid}} \\
    {\color{black}r_{mis}^{pid}} & r_T & {\color{black}r_{mis}^{pid}} \\
    {\color{black}r_{mis}^{pid}} & {\color{black}r_{mis}^{pid}} & r_T 
\end{pmatrix},
\end{equation}
where $r_{T}$ = 1-2${\color{black}r_{mis}^{pid}}$.
The correlation between $\epsilon_{\mathrm{eff}}$ and ${\color{black}r_{mis}^{pid}}$ is illustrated in figure~\ref{fig:ETP_vs_PID}, 

\begin{figure}[htbp]
  \centering
  \includegraphics[width=0.8\textwidth]{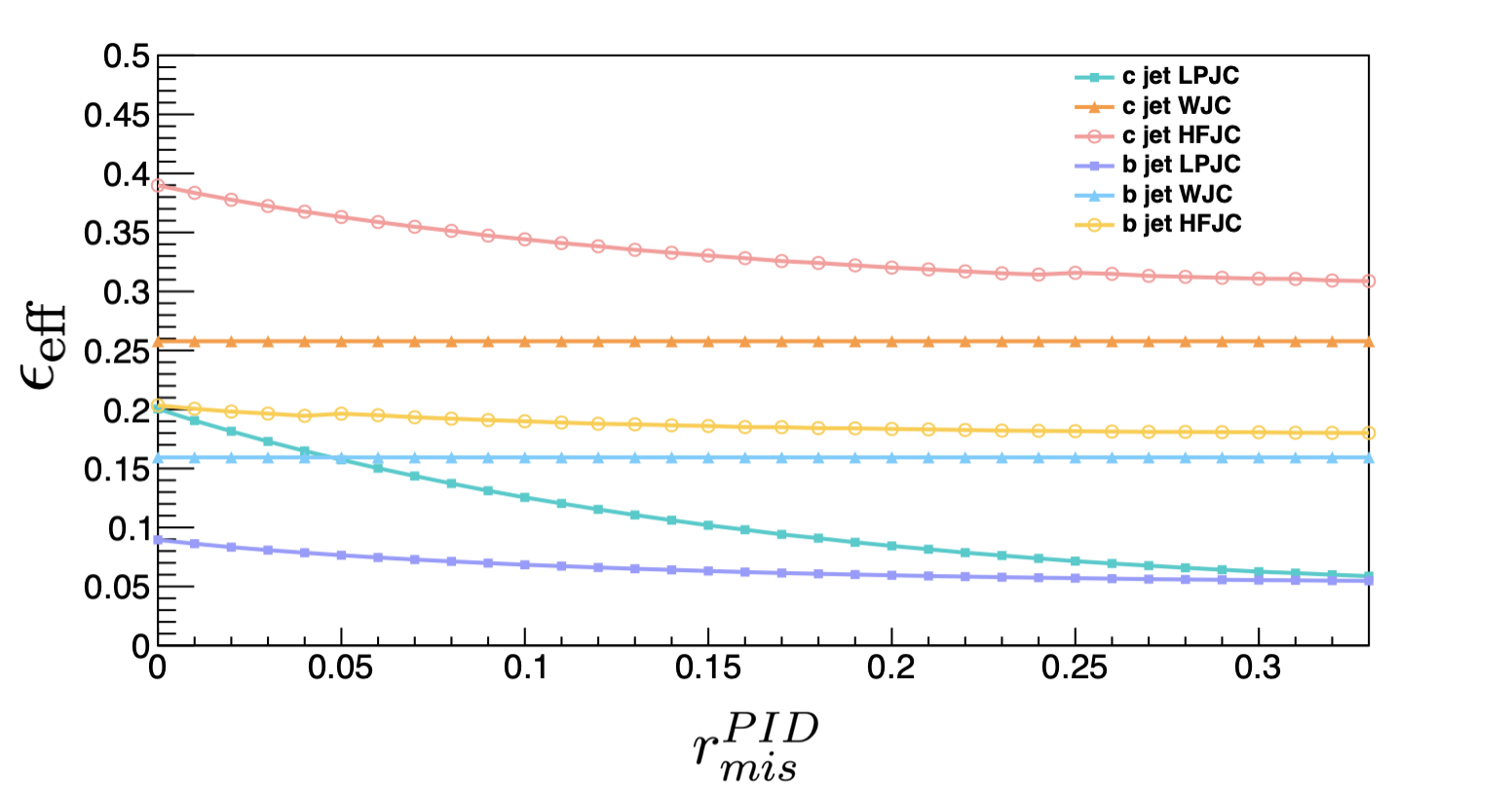}
  \caption{The \etp{} varies with the PID mis-identification rate ${\color{black}r_{mis}^{pid}}$ for \bjet{} and \cjet.}
  \label{fig:ETP_vs_PID}
\end{figure}

The WJC remains unaffected by PID performance as the jet charge calculation involves all final state charged particles, irrespective of their type.
In contrast, the \etp{} of LPJC noticeably diminishes as ${\color{black}r_{mis}^{pid}}$ increases, ultimately reaching an $\epsilon_{\mathrm{eff}}$ of 5.5\% that primarily stems from lepton information.
The \etp{} of \cjet{} decreases faster than that of \bjet{} because leading kaons have different origins and exhibit opposite charge correlation for \cjet.
The PID identification performance at future circular colliders is analyzed in~\cite{An:2018jtk,Zhu:2022hyy}, in which dE/dx resolution is pursued to be better than 3\%, making a 4.6\%/3.9\% degradation of \etp{} for \cbjet, respectively.

\item \textbf {Energy threshold} 

The impact of energy threshold on jet charge \etp{} is illustrated in figure~\ref{fig:ETP_vs_EnergyThreshold}.
The WJC demonstrates greater sensitivity, as the selection efficiency significantly influences the total charge of a jet.
In contrast, the LPJC method exhibits relative insensitivity.
The degradation of \etp{} can be effectively ignored at an energy threshold of 1 \GeV.

\begin{figure}[htbp]
  \centering
  \includegraphics[width=0.8\textwidth]{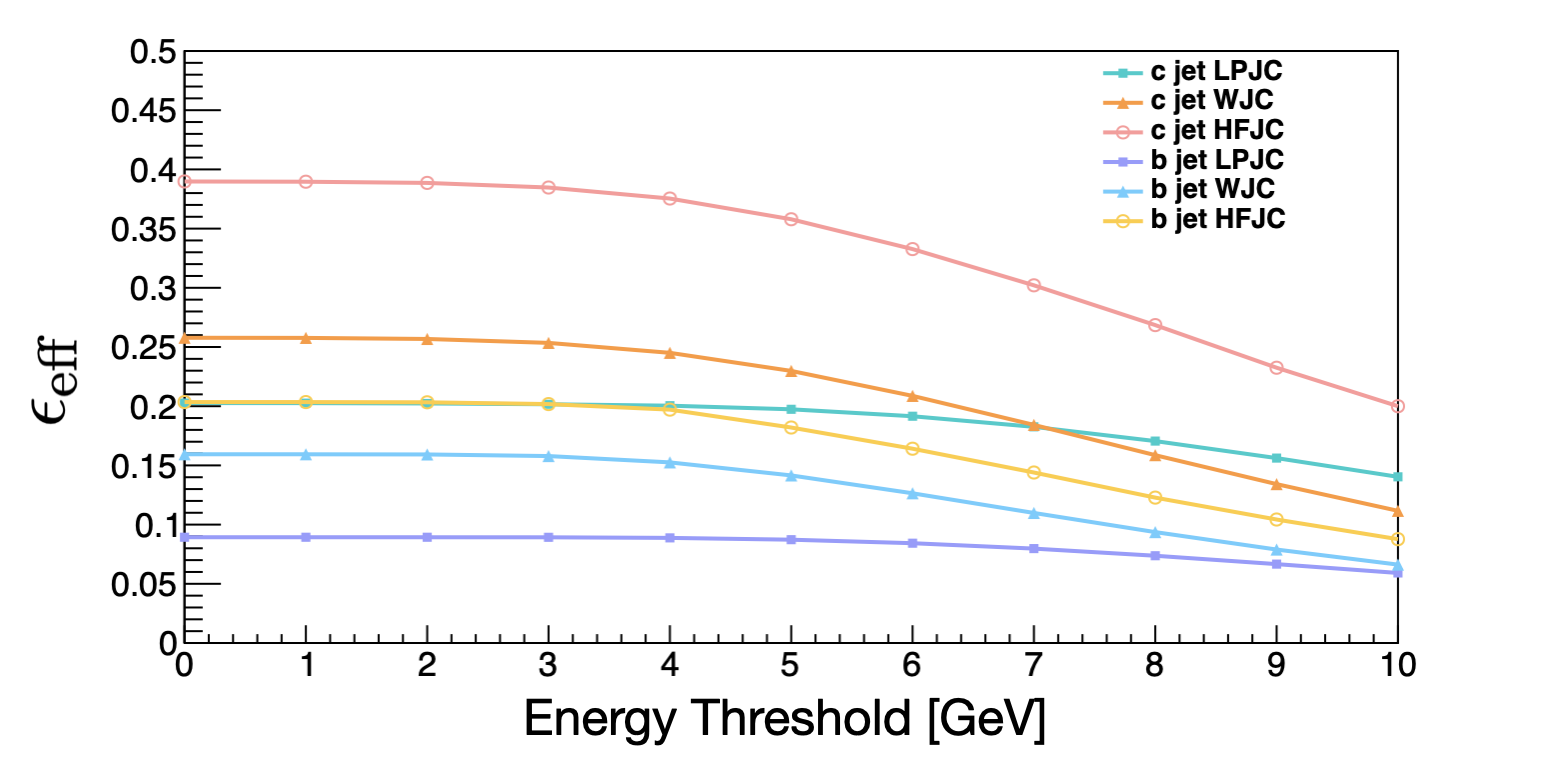}
  \caption{The \etp{} varies with the energy threshold for \bjet{} and \cjet.}
  \label{fig:ETP_vs_EnergyThreshold}
\end{figure}

\item \textbf {Other detector parameters} 

The momentum resolution and polar angle acceptance can be quantified by the Particle Flow Algorithm (PFA)~\cite{Zhao:2017qcy} oriented baseline at Higgs factories, which gives $\delta p/p$ around 0.1\% or better and $|cos(\theta)|$ around 0.99.
The track reconstruction can refer to~\cite{CMS:2012feb}, finding charged particles with transverse momentum \pT{} as low as 0.1 \GeV, or produced as far as 60cm from the beam line.
Under these conditions, the degradation of \etp{} is negligible.

However, there are remaining dilution effects after all effects that we considered or techniques to mitigate dilution.
To correct for these effects, we use the residual dilution, which quantifies the extent to which the measured jet charge deviates from the true jet charge due to these unwanted contributions.
We account for the residual dilution by applying a correction factor to the measured values.
\end{itemize}

\section{Summary}
\label{sec:Summary}

The jet charge plays a crucial role in electroweak and flavor physics measurements at collider experiments. 
In this study, we evaluate the performance of heavy flavor jet charge identifications at future \Z{} factories using truth-level generators. 
The jet charge performance is quantified by the \etp.
We develop a Leading Particle Jet Charge method (LPJC) and combine it with Weighted Jet Charge (WJC) to create a Heavy Flavor Jet Charge method (HFJC), achieving an \etp{} of 39\%/20\% for \cbjet, respectively. 

The effective tagging power is highly dependent on the species of the final state leading charged particle, the species of the leading heavy hadron that \bcquark{} hadronizes into, and the decay source of the final state particles.
Using leading heavy decay products, particularly leptons from charged heavy mesons and protons from baryons(\Lambdac/\Lambdab), LPJC achieves a very low \mromega.
In contrast, the final state leading charged particle from \Bs{} contributes little to the jet charge identification due to the fast oscillation of \Bs.
However, measuring the flavor of the other \bhadron{} produced in the event
is promising to achieve an \etp{} of 20.2\%~\cite{Li:2022tlo}.
The high dependency on leading hadron type indicates a potential for specific channel measurements on jet charge.
If the origin of the final state leading charged particle can be distinguished between leading heavy hadron decay and QCD fragmentation, the \etp{} could be improved to 45\%/37\% for \cbjet.
Furthermore, if each type of leading heavy hadrons can be distinguished, the \etp{} reaches 56.1\%/47.8\% for \cbjet.
The former could, in principle, be approached by an excellent vertex system, while the latter is much more challenging; 
however, carefully analyzing the information of all the final state particles in \Zqq{} event can also provide certain \etp{} for different leading heavy hadron species, especially in specific decay modes.
The values of \etp{} are summarized in Table~\ref{tab:ETP_all_methods}.

\begin{figure}[htbp]
  \centering
  \includegraphics[width=\textwidth]{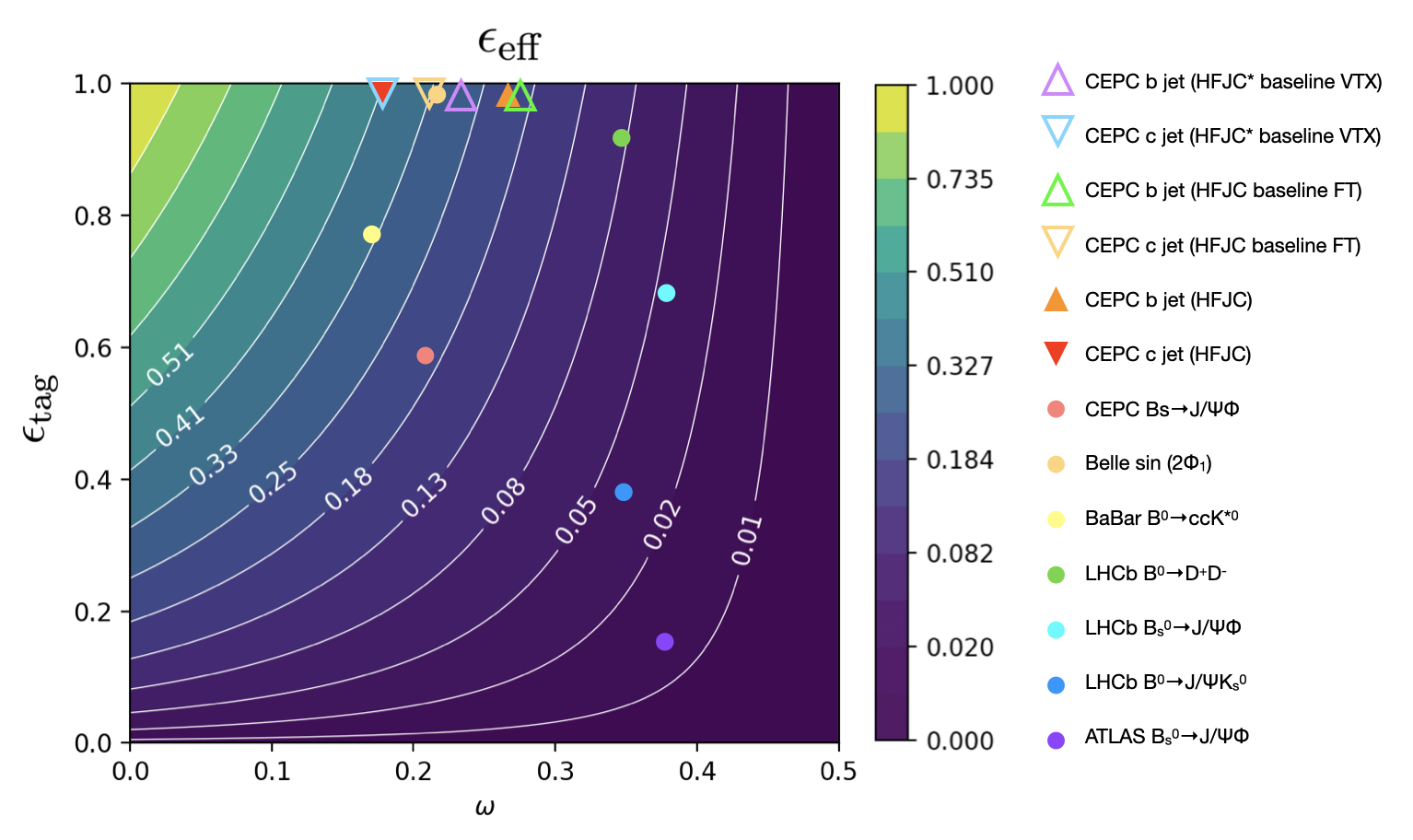}
  \caption{The comparison of jet charge performance between different colliders under different conditions. the color maps the corresponding \etp, the x-axis is the \mromega, and the y-axis refers to the selection \efficiency. The solid triangles emphasize the \etp{} for CEPC employing the HFJC method with an ideal detector, corresponding to the $\epsilon_{\mathrm{tag,max}}$ in Table~\ref{tab:ETP_LPJC}. The hollow triangles labeled "HFJC* Baseline VTX" represent the \etp{} obtained by incorporating the baseline vertex information on top of the solid triangles. The hollow triangles labeled "HFJC Baseline FT" represent the \etp{} achieved using the HFJC method but with baseline flavor tagging efficiency instead of ideal flavor tagging. And the dots represent the results of related jet charge experiments of LHCb, ATLAS, BaBar, and Belle for each specific decay mode in ~\cite{LHCb:2018roe, ATLAS:2016pno} and their corresponding references.}
  \label{fig:compare}
\end{figure}

The impact of critical detector performance on jet charge is also quantified.
On the referenced detector~\cite{CEPCStudyGroup:2018ghi,EuropeanStrategyforParticlePhysicsPreparatoryGroup:2019qin,ECFA2021,Gaudio:2022jve},
most of the effects have a relatively small impact on jet charge performance,
while the PID performance has a more obvious effect on \etp, quantified in figure~\ref{fig:ETP_vs_PID}.
Although our analysis is at the truth level, the reconstructed results are expected to be similar.

{\color{black}A comparison of jet charge performance between colliders under different conditions is displayed in figure~\ref{fig:compare}, 
where 
the color maps the corresponding \etp, 
the white lines indicate contours of constant \etp, 
the x-axis is the \mromega, 
and the y-axis refers to the selection \efficiency{} from all final state charged particles. 
The solid triangles emphasize the \etp{} for CEPC employing HFJC method with an ideal detector, corresponding to the $\epsilon_{\mathrm{tag,max}}$ in Table~\ref{tab:ETP_LPJC}.
The hollow triangles labeled "HFJC* Baseline VTX" represent the \etp{} obtained by incorporating the baseline vertex information on top of the HFJC method represented by solid triangles.
The hollow triangles labeled "HFJC Baseline FT" represent the \etp{} achieved using the HFJC method but with baseline flavor tagging efficiency instead of ideal flavor tagging.
And the dots represent the results of recent jet charge analyses conducted by LHCb, ATLAS, BaBar, and Belle for specific decay modes, leading to a lower efficiency, as reported in the article by~\cite{LHCb:2018roe, ATLAS:2016pno} and its corresponding references.
The results of \Z{} factories are relatively close to the flavor factory case.}

This study provides insights into the factors that affect the correlation between jet charge and fragmentation methods of different generators.
This result has substantial implications for precise electroweak and flavor physics measurements, such as the forward-backward asymmetry \AFB{} and the electroweak mixing angle \swsq~\cite{Yan:2021veo,Zhao:2022lyl,dEnterria:2018jsx}, 
the time-dependent \CP{} measurement~\cite{Chen:2021ftn,dArgent:2021xcs,Bertolin:2021xli,Maccolini:2021rom},
and the Higgs properties measurement at high-energy frontier~\cite{Li:2023tcr,Wong:2023vpx}.
In the future, jet charge identification is expected to be extended to multi-jet events, other center-of-mass energy frontiers, and other collision environments such as $pp$ colliders. 
Besides, investigating detector dependency on the full simulation level is the next logical step.
The jet charge performance versus jet clustering algorithm is also of great interest.
Moreover, the algorithm can be combined with other information, for example, secondary vertex charge, kaon charge for \Bs, neutral kaon identification, jet shape, thrust, and multiplicity information~\cite{Kang:2023ptt}.
Further advancements using machine learning techniques are also worth exploring. 

\appendix

\acknowledgments

We would like to express our gratitude to Gang LI for providing the generator samples used in this study. 
We are also thankful to Lingfeng Li for their valuable suggestion on flavor tagging.
Furthermore, we extend our appreciation to Bin Yan, Yusheng Wu, Dan Yu, and Yongfeng Zhu for their insightful discussions and contributions to this research. 
This project is supported by the International Partnership Program of Chinese Academy of Sciences (Grant No. 113111KYSB20190030), the Innovative Scientific Program of Institute of High Energy Physics.



\bibliography{Jet_Charge_Cuihh.bib}

\providecommand{\href}[2]{#2}\begingroup\raggedright\begin{thebibliography}{10}

\bibitem{HFLAV:2022pwe}
{\bf HFLAV} Collaboration, Y.~S. Amhis et~al., {\it {Averages of $b$-hadron, $c$-hadron, and $\tau$-lepton properties as of 2021}},  \href{http://arxiv.org/abs/2206.07501}{{\tt arXiv:2206.07501}}.

\bibitem{PhysRevD.101.076020}
H.~T. Li and I.~Vitev, {\it Jet charge modification in finite qcd matter},  {\em Phys. Rev. D} {\bf 101} (Apr, 2020) 076020.

\bibitem{Elder:2018owb}
B.~T. Elder, {\em {Jet Fragmentation at the LHC}}.
\newblock PhD thesis, MIT, 2018.

\bibitem{Field:1977fa}
R.~D. Field and R.~P. Feynman, {\it {A Parametrization of the Properties of Quark Jets}},  {\em Nucl. Phys. B} {\bf 136} (1978) 1.

\bibitem{Li:2023tcr}
H.~T. Li, B.~Yan, and C.~P. Yuan, {\it {Discriminating Higgs production mechanisms via jet charge at the LHC}},  \href{http://arxiv.org/abs/2301.07914}{{\tt arXiv:2301.07914}}.

\bibitem{Wong:2023vpx}
X.~Wong and B.~Yan, {\it {Probing the $Hgg$ coupling through the jet charge correlation in Higgs boson decay}},  \href{http://arxiv.org/abs/2302.02084}{{\tt arXiv:2302.02084}}.

\bibitem{ATLAS:2017dfg}
{\bf ATLAS} Collaboration, {\it {Quark versus Gluon Jet Tagging Using Jet Images with the ATLAS Detector}}, .

\bibitem{Elder:2017bkd}
B.~T. Elder, M.~Procura, J.~Thaler, W.~J. Waalewijn, and K.~Zhou, {\it {Generalized Fragmentation Functions for Fractal Jet Observables}},  {\em JHEP} {\bf 06} (2017) 085, [\href{http://arxiv.org/abs/1704.05456}{{\tt arXiv:1704.05456}}].

\bibitem{ATLAS:2017nma}
{\bf ATLAS} Collaboration, {\it {Quark versus Gluon Jet Tagging Using Charged Particle Multiplicity with the ATLAS Detector}}, .

\bibitem{Frye:2017yrw}
C.~Frye, A.~J. Larkoski, J.~Thaler, and K.~Zhou, {\it {Casimir Meets Poisson: Improved Quark/Gluon Discrimination with Counting Observables}},  {\em JHEP} {\bf 09} (2017) 083, [\href{http://arxiv.org/abs/1704.06266}{{\tt arXiv:1704.06266}}].

\bibitem{Larkoski:2014pca}
A.~J. Larkoski, J.~Thaler, and W.~J. Waalewijn, {\it {Gaining (Mutual) Information about Quark/Gluon Discrimination}},  {\em JHEP} {\bf 11} (2014) 129, [\href{http://arxiv.org/abs/1408.3122}{{\tt arXiv:1408.3122}}].

\bibitem{Gallicchio:2011xq}
J.~Gallicchio and M.~D. Schwartz, {\it {Quark and Gluon Tagging at the LHC}},  {\em Phys. Rev. Lett.} {\bf 107} (2011) 172001, [\href{http://arxiv.org/abs/1106.3076}{{\tt arXiv:1106.3076}}].

\bibitem{Lee:2022kdn}
K.~Lee, J.~Mulligan, M.~P\l{}osko\'n, F.~Ringer, and F.~Yuan, {\it {Machine learning-based jet and event classification at the Electron-Ion Collider with applications to hadron structure and spin physics}},  \href{http://arxiv.org/abs/2210.06450}{{\tt arXiv:2210.06450}}.

\bibitem{Ilten:2017rbd}
P.~Ilten, N.~L. Rodd, J.~Thaler, and M.~Williams, {\it {Disentangling Heavy Flavor at Colliders}},  {\em Phys. Rev. D} {\bf 96} (2017), no.~5 054019, [\href{http://arxiv.org/abs/1702.02947}{{\tt arXiv:1702.02947}}].

\bibitem{CMS:2017yer}
{\bf CMS} Collaboration, A.~M. Sirunyan et~al., {\it {Measurements of jet charge with dijet events in pp collisions at $\sqrt{s}=8$ TeV}},  {\em JHEP} {\bf 10} (2017) 131, [\href{http://arxiv.org/abs/1706.05868}{{\tt arXiv:1706.05868}}].

\bibitem{Yan:2021veo}
B.~Yan and C.~P. Yuan, {\it {Anomalous Zbb\textasciimacron{} Couplings: From LEP to LHC}},  {\em Phys. Rev. Lett.} {\bf 127} (2021), no.~5 051801, [\href{http://arxiv.org/abs/2101.06261}{{\tt arXiv:2101.06261}}].

\bibitem{Zhao:2022lyl}
Z.~Zhao, S.~Yang, M.~Ruan, M.~Liu, and L.~Han, {\it {Measurement of the effective weak mixing angle at the CEPC}},  \href{http://arxiv.org/abs/2204.09921}{{\tt arXiv:2204.09921}}.

\bibitem{dEnterria:2018jsx}
D.~d'Enterria and C.~Yan, {\it {Forward-backward $b$-quark asymmetry at the Z pole: QCD uncertainties redux}},  in {\em {53rd Rencontres de Moriond on QCD and High Energy Interactions}}, pp.~253--257, 2018.
\newblock \href{http://arxiv.org/abs/1806.00141}{{\tt arXiv:1806.00141}}.

\bibitem{DELPHI:2004wzo}
{\bf DELPHI} Collaboration, J.~Abdallah et~al., {\it {Determination of A**b(FB) at the Z pole using inclusive charge reconstruction and lifetime tagging}},  {\em Eur. Phys. J. C} {\bf 40} (2005) 1--25, [\href{http://arxiv.org/abs/hep-ex/0412004}{{\tt hep-ex/0412004}}].

\bibitem{OPAL:2003pfe}
{\bf OPAL} Collaboration, G.~Abbiendi et~al., {\it {Measurement of heavy quark forward backward asymmetries and average B mixing using leptons in hadronic Z decays}},  {\em Phys. Lett. B} {\bf 577} (2003) 18--36, [\href{http://arxiv.org/abs/hep-ex/0308051}{{\tt hep-ex/0308051}}].

\bibitem{ALEPH:2001pzx}
{\bf ALEPH} Collaboration, A.~Heister et~al., {\it {Measurement of the forward backward asymmetry in Z --\ensuremath{>} b anti-b and Z --\ensuremath{>} c anti-c decays with leptons}},  {\em Eur. Phys. J. C} {\bf 24} (2002) 177--191.

\bibitem{OPAL:2002vqi}
{\bf OPAL} Collaboration, G.~Abbiendi et~al., {\it {Measurement of the b quark forward backward asymmetry around the Z0 peak using an inclusive tag}},  {\em Phys. Lett. B} {\bf 546} (2002) 29--47, [\href{http://arxiv.org/abs/hep-ex/0209076}{{\tt hep-ex/0209076}}].

\bibitem{ALEPH:2001mdb}
{\bf ALEPH} Collaboration, A.~Heister et~al., {\it {Measurement of A**b(FB) using inclusive b hadron decays}},  {\em Eur. Phys. J. C} {\bf 22} (2001) 201--215, [\href{http://arxiv.org/abs/hep-ex/0107033}{{\tt hep-ex/0107033}}].

\bibitem{DELPHI:1994yxx}
{\bf DELPHI} Collaboration, P.~Abreu et~al., {\it {Measurement of the forward - backward asymmetry of e+ e- ---\ensuremath{>} Z ---\ensuremath{>} b anti-b using prompt leptons and a lifetime tag}},  {\em Z. Phys. C} {\bf 65} (1995) 569--586.

\bibitem{L3:1992fsb}
{\bf L3} Collaboration, O.~Adriani et~al., {\it {Measurement of the e+ e- --\ensuremath{>} b anti-b and e+ e- --\ensuremath{>} c anti-c forward backward asymmetries at the Z0 resonance}},  {\em Phys. Lett. B} {\bf 292} (1992) 454--462.

\bibitem{L3:1998jgx}
{\bf L3} Collaboration, M.~Acciarri et~al., {\it {Measurement of the effective weak mixing angle by jet charge asymmetry in hadronic decays of the Z boson}},  {\em Phys. Lett. B} {\bf 439} (1998) 225--236.

\bibitem{Schael:1991nf}
S.~Schael, {\it {Measurements of sin**2 theta-W from the charge asymmetry of hadronic events at the Z0 peak}},  in {\em {Joint International Lepton Photon Symposium at High Energies (15th) and European Physical Society Conference on High-energy Physics}}, 11, 1991.

\bibitem{Chen:2021ftn}
S.~Chen, Y.~Li, W.~Qian, Y.~Xie, Z.~Yang, L.~Zhang, and Y.~Zhang, {\it {Heavy Flavour Physics and CP Violation at LHCb: a Ten-Year Review}},  \href{http://arxiv.org/abs/2111.14360}{{\tt arXiv:2111.14360}}.

\bibitem{dArgent:2021xcs}
{\bf LHCb} Collaboration, P.~d'Argent, {\it {Beautiful mixing and CP violation at LHCb}},  in {\em {55th Rencontres de Moriond on QCD and High Energy Interactions}}, 11, 2021.
\newblock \href{http://arxiv.org/abs/2111.06786}{{\tt arXiv:2111.06786}}.

\bibitem{Bertolin:2021xli}
A.~Bertolin, {\it {CP violation in beauty with the LHCb experiment}},  {\em PoS} {\bf LHCP2021} (2021) 049.

\bibitem{Maccolini:2021rom}
S.~Maccolini, {\it {CP violation in charm with the LHCb experiment}},  {\em PoS} {\bf LHCP2021} (2021) 048.

\bibitem{Heinicke:2229990}
K.~Heinicke, {\it {Optimization of Flavour Tagging Algorithms for the LHCb Experiment}},  Sep, 2016.
\newblock Presented 30 Sep 2016.

\bibitem{Belle-II:2018jsg}
{\bf Belle-II} Collaboration, W.~Altmannshofer et~al., {\it {The Belle II Physics Book}},  {\em PTEP} {\bf 2019} (2019), no.~12 123C01, [\href{http://arxiv.org/abs/1808.10567}{{\tt arXiv:1808.10567}}]. [Erratum: PTEP 2020, 029201 (2020)].

\bibitem{CDF:1999jfn}
{\bf CDF} Collaboration, F.~Abe et~al., {\it {Measurement of $B^0 \overline{B}^0$ flavor oscillations using jet-charge and lepton flavor tagging in $p\bar{p}$ collisions at $\sqrt{s} = 1.8$ TeV}},  {\em Phys. Rev. D} {\bf 60} (1999) 072003, [\href{http://arxiv.org/abs/hep-ex/9903011}{{\tt hep-ex/9903011}}].

\bibitem{LEPHeavyFlavorWorkingGroup:1998roh}
{\bf LEP Heavy Flavor Working Group} Collaboration, D.~Abbaneo, P.~Antilogus, T.~Behnke, S.~C. Blyth, M.~Elsing, R.~Faccini, R.~W.~L. Jones, K.~Monig, S.~Petzold, and R.~Tenchini, {\it {QCD corrections to the forward - backward asymmetries of c and b quarks at the Z pole}},  {\em Eur. Phys. J. C} {\bf 4} (1998) 185--191.

\bibitem{OPAL:1994xvz}
{\bf OPAL} Collaboration, R.~Akers et~al., {\it {Measurement of the time dependence of B(d)0 \ensuremath{<}---\ensuremath{>} anti-B(d)0 mixing using a jet charge technique}},  {\em Phys. Lett. B} {\bf 327} (1994) 411--424.

\bibitem{ALEPH:1991fba}
{\bf ALEPH} Collaboration, D.~Decamp et~al., {\it {Measurement of charge asymmetry in hadronic Z decays}},  {\em Phys. Lett. B} {\bf 259} (1991) 377--388.

\bibitem{Bielcikova:2021ujk}
J.~Bielcikova, {\it {Summary: Jets and High-p$_T$}},  {\em PoS} {\bf HardProbes2020} (2021) 028.

\bibitem{CMS:2020plq}
{\bf CMS} Collaboration, A.~M. Sirunyan et~al., {\it {Measurement of quark- and gluon-like jet fractions using jet charge in PbPb and pp collisions at 5.02 TeV}},  {\em JHEP} {\bf 07} (2020) 115, [\href{http://arxiv.org/abs/2004.00602}{{\tt arXiv:2004.00602}}].

\bibitem{ATLAS:2015rlw}
{\bf ATLAS} Collaboration, G.~Aad et~al., {\it {Measurement of jet charge in dijet events from $\sqrt{s}$=8 TeV pp collisions with the ATLAS detector}},  {\em Phys. Rev. D} {\bf 93} (2016), no.~5 052003, [\href{http://arxiv.org/abs/1509.05190}{{\tt arXiv:1509.05190}}].

\bibitem{Nachman:2014qma}
{\bf ATLAS} Collaboration, B.~Nachman, {\it {Jet Charge with the ATLAS Detector using $\sqrt{s}=8$ TeV $pp$ Collision Data}},  in {\em {2nd Large Hadron Collider Physics Conference}}, 9, 2014.
\newblock \href{http://arxiv.org/abs/1409.0318}{{\tt arXiv:1409.0318}}.

\bibitem{Krohn:2012fg}
D.~Krohn, M.~D. Schwartz, T.~Lin, and W.~J. Waalewijn, {\it {Jet Charge at the LHC}},  {\em Phys. Rev. Lett.} {\bf 110} (2013), no.~21 212001, [\href{http://arxiv.org/abs/1209.2421}{{\tt arXiv:1209.2421}}].

\bibitem{ATLAS:2016pno}
{\bf ATLAS} Collaboration, G.~Aad et~al., {\it {Measurement of the CP-violating phase $\phi_s$ and the $B^0_s$ meson decay width difference with $B^0_s \to J/\psi\phi$ decays in ATLAS}},  {\em JHEP} {\bf 08} (2016) 147, [\href{http://arxiv.org/abs/1601.03297}{{\tt arXiv:1601.03297}}].

\bibitem{LHCb:2018roe}
{\bf LHCb} Collaboration, R.~Aaij et~al., {\it {Physics case for an LHCb Upgrade II - Opportunities in flavour physics, and beyond, in the HL-LHC era}},  \href{http://arxiv.org/abs/1808.08865}{{\tt arXiv:1808.08865}}.

\bibitem{Gianelle:2022unu}
A.~Gianelle, P.~Koppenburg, D.~Lucchesi, D.~Nicotra, E.~Rodrigues, L.~Sestini, J.~de~Vries, and D.~Zuliani, {\it {Quantum Machine Learning for b-jet charge identification}},  {\em JHEP} {\bf 08} (2022) 014, [\href{http://arxiv.org/abs/2202.13943}{{\tt arXiv:2202.13943}}].

\bibitem{BaBar:2009byl}
{\bf BaBar} Collaboration, B.~Aubert et~al., {\it {Measurement of Time-Dependent CP Asymmetry in B0 ---\ensuremath{>} c anti-c K(*)0 Decays}},  {\em Phys. Rev. D} {\bf 79} (2009) 072009, [\href{http://arxiv.org/abs/0902.1708}{{\tt arXiv:0902.1708}}].

\bibitem{BaBar:2012fgk}
{\bf BaBar} Collaboration, J.~P. Lees et~al., {\it {Measurement of CP Asymmetries and Branching Fractions in Charmless Two-Body $B$-Meson Decays to Pions and Kaons}},  {\em Phys. Rev. D} {\bf 87} (2013), no.~5 052009, [\href{http://arxiv.org/abs/1206.3525}{{\tt arXiv:1206.3525}}].

\bibitem{Belle-II:2021zvj}
{\bf Belle-II} Collaboration, F.~Abudin\'en et~al., {\it {B-flavor tagging at Belle II}},  {\em Eur. Phys. J. C} {\bf 82} (2022), no.~4 283, [\href{http://arxiv.org/abs/2110.00790}{{\tt arXiv:2110.00790}}].

\bibitem{EuropeanStrategyforParticlePhysicsPreparatoryGroup:2019qin}
R.~K. Ellis et~al., {\it {Physics Briefing Book}: {Input for the European Strategy for Particle Physics Update 2020}},  \href{http://arxiv.org/abs/1910.11775}{{\tt arXiv:1910.11775}}.

\bibitem{CEPCStudyGroup:2018rmc}
{\bf CEPC Study Group} Collaboration, {\it {CEPC Conceptual Design Report: Volume 1 - Accelerator}},  \href{http://arxiv.org/abs/1809.00285}{{\tt arXiv:1809.00285}}.

\bibitem{CEPCStudyGroup:2018ghi}
{\bf CEPC Study Group} Collaboration, M.~Dong et~al., {\it {CEPC Conceptual Design Report: Volume 2 - Physics \& Detector}},  \href{http://arxiv.org/abs/1811.10545}{{\tt arXiv:1811.10545}}.

\bibitem{FCC:2018byv}
{\bf FCC} Collaboration, A.~Abada et~al., {\it {FCC Physics Opportunities}: {Future Circular Collider Conceptual Design Report Volume 1}},  {\em Eur. Phys. J. C} {\bf 79} (2019), no.~6 474.

\bibitem{FCC:2018evy}
{\bf FCC} Collaboration, A.~Abada et~al., {\it {FCC-ee: The Lepton Collider}: {Future Circular Collider Conceptual Design Report Volume 2}},  {\em Eur. Phys. J. ST} {\bf 228} (2019), no.~2 261--623.

\bibitem{FCC:2018vvp}
{\bf FCC} Collaboration, A.~Abada et~al., {\it {FCC-hh: The Hadron Collider}: {Future Circular Collider Conceptual Design Report Volume 3}},  {\em Eur. Phys. J. ST} {\bf 228} (2019), no.~4 755--1107.

\bibitem{ILC:2013jhg}
{\bf ILC} Collaboration, {\it {The International Linear Collider Technical Design Report - Volume 2: Physics}},  \href{http://arxiv.org/abs/1306.6352}{{\tt arXiv:1306.6352}}.

\bibitem{Linssen:2012hp}
{\it {Physics and Detectors at CLIC: CLIC Conceptual Design Report}},  \href{http://arxiv.org/abs/1202.5940}{{\tt arXiv:1202.5940}}.

\bibitem{Bai:2021rdg}
M.~Bai et~al., {\it {C$^3$: A ''Cool'' Route to the Higgs Boson and Beyond}},  in {\em {Snowmass 2021}}, 10, 2021.
\newblock \href{http://arxiv.org/abs/2110.15800}{{\tt arXiv:2110.15800}}.

\bibitem{Belomestnykh:2022wbn}
S.~Belomestnykh et~al., {\it {Higgs-Energy LEptoN (HELEN) Collider based on advanced superconducting radio frequency technology}},  in {\em {Snowmass 2021}}, 3, 2022.
\newblock \href{http://arxiv.org/abs/2203.08211}{{\tt arXiv:2203.08211}}.

\bibitem{Zhu:2023xpk}
Y.~Zhu, H.~Liang, Y.~Wang, H.~Qu, C.~Zhou, and M.~Ruan, {\it {ParticleNet and its application on CEPC Jet Flavor Tagging}},  \href{http://arxiv.org/abs/2309.13231}{{\tt arXiv:2309.13231}}.

\bibitem{Stienemeier:2021cse}
P.~Stienemeier, S.~Bra\ss{}, P.~Bredt, W.~Kilian, N.~Kreher, T.~Ohl, J.~Reuter, V.~Rothe, and T.~Striegl, {\it {WHIZARD 3.0: Status and News}},  in {\em {International Workshop on Future Linear Colliders}}, 4, 2021.
\newblock \href{http://arxiv.org/abs/2104.11141}{{\tt arXiv:2104.11141}}.

\bibitem{Brass:2019hvu}
S.~Bra\ss{}, W.~Kilian, T.~Ohl, J.~Reuter, V.~Rothe, and P.~Stienemeier, {\it {Precision Monte Carlo simulations with WHIZARD}},  {\em CERN Yellow Reports: Monographs} {\bf 3} (2020) 205--210.

\bibitem{Kilian:2007gr}
W.~Kilian, T.~Ohl, and J.~Reuter, {\it {WHIZARD: Simulating Multi-Particle Processes at LHC and ILC}},  {\em Eur. Phys. J. C} {\bf 71} (2011) 1742, [\href{http://arxiv.org/abs/0708.4233}{{\tt arXiv:0708.4233}}].

\bibitem{Bellm:2015jjp}
J.~Bellm et~al., {\it {Herwig 7.0/Herwig++ 3.0 release note}},  {\em Eur. Phys. J. C} {\bf 76} (2016), no.~4 196, [\href{http://arxiv.org/abs/1512.01178}{{\tt arXiv:1512.01178}}].

\bibitem{Sherpa:2019gpd}
{\bf Sherpa} Collaboration, E.~Bothmann et~al., {\it {Event Generation with Sherpa 2.2}},  {\em SciPost Phys.} {\bf 7} (2019), no.~3 034, [\href{http://arxiv.org/abs/1905.09127}{{\tt arXiv:1905.09127}}].

\bibitem{Sjostrand:2006za}
T.~Sjostrand, S.~Mrenna, and P.~Z. Skands, {\it {PYTHIA 6.4 Physics and Manual}},  {\em JHEP} {\bf 05} (2006) 026, [\href{http://arxiv.org/abs/hep-ph/0603175}{{\tt hep-ph/0603175}}].

\bibitem{Waalewijn:2012sv}
W.~J. Waalewijn, {\it {Calculating the Charge of a Jet}},  {\em Phys. Rev. D} {\bf 86} (2012) 094030, [\href{http://arxiv.org/abs/1209.3019}{{\tt arXiv:1209.3019}}].

\bibitem{ParticleDataGroup:2020ssz}
{\bf Particle Data Group} Collaboration, P.~A. Zyla et~al., {\it {Review of Particle Physics}},  {\em PTEP} {\bf 2020} (2020), no.~8 083C01.

\bibitem{Chang:2013iba}
H.-M. Chang, M.~Procura, J.~Thaler, and W.~J. Waalewijn, {\it {Calculating Track Thrust with Track Functions}},  {\em Phys. Rev. D} {\bf 88} (2013) 034030, [\href{http://arxiv.org/abs/1306.6630}{{\tt arXiv:1306.6630}}].

\bibitem{LHCb:2021moh}
{\bf LHCb} Collaboration, R.~Aaij et~al., {\it {Precise determination of the $B^0_s$-$\overline{B}^0_s$ oscillation frequency}},  \href{http://arxiv.org/abs/2104.04421}{{\tt arXiv:2104.04421}}.

\bibitem{Lenz:2006hd}
A.~Lenz and U.~Nierste, {\it {Theoretical update of $B_s - \bar{B}_s$ mixing}},  {\em JHEP} {\bf 06} (2007) 072, [\href{http://arxiv.org/abs/hep-ph/0612167}{{\tt hep-ph/0612167}}].

\bibitem{CDF:2006imy}
{\bf CDF} Collaboration, A.~Abulencia et~al., {\it {Observation of $B^0_s - \bar{B}^0_s$ Oscillations}},  {\em Phys. Rev. Lett.} {\bf 97} (2006) 242003, [\href{http://arxiv.org/abs/hep-ex/0609040}{{\tt hep-ex/0609040}}].

\bibitem{LHCb:2013lrq}
{\bf LHCb} Collaboration, R.~Aaij et~al., {\it {Precision measurement of the $B^{0}_{s}$-$\bar{B}^{0}_{s}$ oscillation frequency with the decay $B^{0}_{s}\rightarrow D^{-}_{s}\pi^{+}$}},  {\em New J. Phys.} {\bf 15} (2013) 053021, [\href{http://arxiv.org/abs/1304.4741}{{\tt arXiv:1304.4741}}].

\bibitem{ALEPH:1992net}
{\bf ALEPH} Collaboration, D.~Buskulic et~al., {\it {Measurement of B - anti-B mixing at the Z using a jet charge method}},  {\em Phys. Lett. B} {\bf 284} (1992) 177--190.

\bibitem{Carrasco:2014nda}
N.~Carrasco, {\it {Neutral meson oscillations on the lattice}},  {\em Nucl. Part. Phys. Proc.} {\bf 273-275} (2016) 1631--1637, [\href{http://arxiv.org/abs/1410.0161}{{\tt arXiv:1410.0161}}].

\bibitem{Li:2022tlo}
X.~Li, M.~Ruan, and M.~Zhao, {\it {Prospect for measurement of CP-violation phase $\phi_s$ study in the $B_s\rightarrow J/\Psi\phi$ channel at future $Z$ factory}},  \href{http://arxiv.org/abs/2205.10565}{{\tt arXiv:2205.10565}}.

\bibitem{ECFA2021}
E.~Collaboration, {\it The 2021 ecfa detector research and development roadmap},  {\em Journal of Instrumentation} {\bf 16} (2021) T06001.

\bibitem{Gaudio:2022jve}
G.~Gaudio, {\it {The IDEA detector concept for FCCee}},  {\em PoS} {\bf ICHEP2022} (11, 2022) 337.

\bibitem{Yu:2021pxc}
D.~Yu, T.~Zheng, and M.~Ruan, {\it {Lepton identification performance in Jets at a future electron positron Higgs Z factory}},  \href{http://arxiv.org/abs/2105.01246}{{\tt arXiv:2105.01246}}.

\bibitem{An:2018jtk}
F.~An, S.~Prell, C.~Chen, J.~Cochran, X.~Lou, and M.~Ruan, {\it {Monte Carlo study of particle identification at the CEPC using TPC dE / dx information}},  {\em Eur. Phys. J. C} {\bf 78} (2018), no.~6 464, [\href{http://arxiv.org/abs/1803.05134}{{\tt arXiv:1803.05134}}].

\bibitem{Zhu:2022hyy}
Y.~Zhu, S.~Chen, H.~Cui, and M.~Ruan, {\it {Requirement analysis for dE/dx measurement and PID performance at the CEPC baseline detector}},  {\em Nucl. Instrum. Meth. A} {\bf 1047} (2023) 167835, [\href{http://arxiv.org/abs/2209.14486}{{\tt arXiv:2209.14486}}].

\bibitem{Zhao:2017qcy}
H.~Zhao, C.~Fu, D.~Yu, Z.~Wang, T.~Hu, and M.~Ruan, {\it {Particle flow oriented electromagnetic calorimeter optimization for the circular electron positron collider}},  {\em JINST} {\bf 13} (2018), no.~03 P03010, [\href{http://arxiv.org/abs/1712.09625}{{\tt arXiv:1712.09625}}].

\bibitem{CMS:2012feb}
{\bf CMS} Collaboration, S.~Chatrchyan et~al., {\it {Identification of b-Quark Jets with the CMS Experiment}},  {\em JINST} {\bf 8} (2013) P04013, [\href{http://arxiv.org/abs/1211.4462}{{\tt arXiv:1211.4462}}].

\bibitem{Kang:2023ptt}
Z.-B. Kang, A.~J. Larkoski, and J.~Yang, {\it {Understanding Jet Charge}},  \href{http://arxiv.org/abs/2301.09649}{{\tt arXiv:2301.09649}}.

\end{thebibliography}\endgroup
\bibliographystyle{JHEP}

\end{document}